\newif\ifjournal
  \journaltrue

\ifjournal
  \documentclass[]{aa}
  \usepackage{graphicx,amssymb,natbib,times,mathptmx}
  \renewcommand{\d}{\mathrm{d}}
  \authorrunning{C. Fedeli et al.}
  \titlerunning
   {Computing strong-lensing cross sections of
    merging clusters}
\else
  \documentclass{paper}
\fi

\begin{document}

\title
  {A fast method for computing strong-lensing cross sections:
   Application to merging clusters}

\ifjournal
\author{Cosimo Fedeli\inst{1,3}, Massimo Meneghetti\inst{1}, Matthias
  Bartelmann\inst{1}, Klaus Dolag\inst{2} and Lauro Moscardini\inst{3}
  \institute
  {$^1$ Zentrum f\"ur Astronomie, ITA, Universit\"at Heidelberg,
   Albert-\"Uberle-Str. 2, 69120 Heidelberg, Germany\\
   $^2$ Max-Planck-Institut f\"ur Astrophysik, Karl-Schwarzschild-Str. 1,
   85740 Garching, Germany\\ 
   $^3$ Dipartimento di Astronomia, Universit\`a di Bologna, via
   Ranzani 1, 40127 Bologna, Italy}}
\else
\author{Cosimo Fedeli$^1$, Massimo Meneghetti$^1$, Matthias
 Bartelmann$^1$, Klaus Dolag$^2$ and Lauro Moscardini$^3$\\
 $^1$ Zentrum f\"ur Astronomie, ITA, Universit\"at Heidelberg,
 Albert-\"Uberle-Str. 2, 69120 Heidelberg, Germany\\
 $^2$ Max-Planck-Institut f\"ur Astrophysik, Karl-Schwarzschild-Str. 1,
 85740 Garching, Germany\\ 
 $^3$ Dipartimento di Astronomia, Universit\`a di Bologna, via
 Ranzani 1, 40127 Bologna, Italy}
\fi

\date{\emph{Astronomy \& Astrophysics, submitted}}

\newcommand{\abstext}
 {Strong gravitational lensing by irregular mass distributions,
  such as galaxy clusters, is generally not well quantified by cross
  sections of analytic mass models. Computationally expensive
  ray-tracing methods have so far been necessary for accurate
  cross-section calculations.
  We describe a fast, semi-analytic method here which is based on
  surface integrals over high-magnification regions in the lens plane and
  demonstrate that it yields reliable cross sections even for complex,
  asymmetric mass distributions. The method is faster than ray-tracing
  simulations by factors of $\sim30$ and thus suitable for large
  cosmological simulations, saving large amounts of computing time.
  We apply this method to a sample of galaxy
  cluster-sized dark matter haloes
  with simulated merger trees and show that cluster mergers
  approximately double the strong-lensing optical depth for
  lens redshifts $z_\mathrm{l}\gtrsim0.5$ and sources
  near $z_\mathrm{s} = 2$. We believe
  that this result hints at one possibility for understanding
  the recently detected high arcs abundance in clusters at moderate 
  and high redshifts, and is thus worth further studies.}

\ifjournal
  \abstract{\abstext}
\else
  \begin{abstract}\abstext\end{abstract}
\fi


\maketitle

\section{Introduction}

Strong gravitational lensing by galaxy clusters is a highly non-linear
effect which is very sensitive to the details of the lensing mass
distribution. The cluster core densities, the asymmetries of their
mass distribution, their substructures and their close neighbourhood
all contribute to their lensing properties. The ongoing debate about
whether the observed statistics of arcs is or is not compatible with
the expectations in the standard $\Lambda$CDM cosmology shows that it
is not sufficiently understood yet what aspects of the source and
cluster populations as a whole determines the statistics of its strong-lensing
effects
\citep{BA98.2,ME00.1,ME03.1,ME03.2,WA03.1,DA03.1,LI05.1,HE05.1}.

It is an obstacle for theoretical as well as observational studies
that cross sections of galaxy clusters for strong lensing are costly
to compute. So far, they require highly-resolved simulations tracing
large numbers of light rays through realistically simulated cluster
mass distributions, used for finding the images of sources which need
to be classified automatically \citep{MI93.1,BA94.1}. This needs to be
done often, i.e.~for different cosmological models and
for many clusters in large cosmological volumes seen
under many different angles, for the results to reach a reliable
level. The recent finding that the enhanced tidal field around merging
clusters substantially enhances strong-lensing cross sections
\citep{TO04.1} adds the necessity to study clusters with a time
resolution which is sufficiently fine to resolve cluster merger
events.

The increasing demands to be met by reliable strong-lensing
calculations and the wish to carry them out for varying cosmological
models calls for a substantially faster method than ray-tracing which
should be equally reliable. We develop such a method here. It is based
on the fact that highly elongated arcs
occur near the critical curves in the lens plane, and that
imaging properties near critical curves can be summarised by the
eigenvalues of the Jacobian matrix of the lens mapping (see \cite{SC92.1}
and Sect.~2 for detail). This allows
the reduction of the cross-section calculation to an area integral to
be carried out on the lens plane itself. In that sense, the method is
analytic, but the irregular shapes of the integration domains require
it to be carried out numerically. Since the eigenvalues of the
Jacobian matrix ideally describe imaging properties for
infinitesimally small, circular sources, the method needs to be
supplemented by techniques for taking extended, elliptical sources
into account without losing computational speed.

The paper describes this method and its testing. After a brief summary
of gravitational lensing in Sect.~2, we develop the new method and
compare it to ray-tracing techniques in Sect.~3. We then describe in
Sect.~4 its application to samples of cluster-sized halos whose
evolution is described by merger trees. Results on cluster cross
sections and optical depths are given in Sect.~5, and Sect.~6 provides
summary and discussion.

\section{Basic Gravitational Lensing}

We briefly compile the basic equations of gravitational lensing
necessary for quantifying the statistics of strong lensing
events. Adopting the thin-lens approximation, the lensing mass
distribution is projected perpendicular to the line of sight onto the
lens plane, on which the physical coordinates $\vec\xi=(\xi_1,\xi_2)$
are introduced. The coordinates on the source plane are
$\vec\eta=(\eta_1,\eta_2)$. Throughout this paper, we shall assume for
simplicity that all sources are at the same redshift because their
distribution in redshift is not relevant for developing our method. By
means of the length scales $\xi_0$ and
$\eta_0=(D_\mathrm{s}/D_\mathrm{l})\xi_0$ in the lens and source
planes, respectively, we can introduce the dimension-less coordinates
$\vec x=\vec\xi/\xi_0$ and $\vec y=\vec\eta/\eta_0$. The
angular-diameter distances from the observer to the lens, the source,
and from the lens to the source are $D_\mathrm{l,s,ls}$, respectively.

Source and image coordinates are related by the lens equation,
\begin{equation}
  \vec y=\vec x-\vec\nabla\psi(\vec x)=\vec x-\vec\alpha(\vec x)\;,
\label{eqn:lens}
\end{equation}
where $\vec\alpha(\vec x)$ is the deflection angle, which is the
gradient of the scalar lensing potential $\psi(\vec x)$ (see,
e.g.~\citealt{SC92.1}). The Poisson equation
\begin{equation}
  \Delta\psi(\vec x)=2\kappa(\vec x)=
  2\frac{\Sigma(\vec x)}{\Sigma_\mathrm{cr}}
\label{eqn:poisson}
\end{equation}
relates the lensing potential to the (dimension-less) convergence
$\kappa$, which is the surface-mass density $\Sigma$ in units of its
critical value $\Sigma_\mathrm{cr}$.


The lens equation (\ref{eqn:lens}) defines a mapping from the lens to
the source plane whose Jacobian matrix $\mathcal{A}(\vec x)$ describes
how the shape of an infinitesimal source located at position $\vec y$
is changed by the lens. It can be decomposed as
\begin{equation}
  \mathcal{A}=\left(\begin{array}{cc}
  1-\kappa-\gamma_1 & -\gamma_2\\
  -\gamma_2 & 1-\kappa+\gamma_1\\
  \end{array}\right)\;,
\label{eqn:matrix}
\end{equation}
where the shear components $\gamma_{1,2}$ form a trace-less tensor
quantifying the lensing distortion. The magnification of an image
relative to the source is given by the inverse Jacobian determinant,
\begin{equation}
  \mu(\vec x)=\frac{1}{\det\mathcal{A}(\vec x)}=
  \frac{1}{[1-\kappa(\vec x)]^2-
              \gamma_1^2(\vec x)-\gamma_2^2(\vec x)}\;.
\label{eqn:magnification}
\end{equation}
Since the deflection angle is a gradient field, $\mathcal{A}(\vec x)$
is symmetric at any point on the lens plane (at least in the single-plane
approximation which we are considering here), thus a coordinate
rotation can always be found which diagonalises $\mathcal{A}$. Its two
eigenvalues
\begin{equation}
  \lambda_\mathrm{t}\equiv1-\kappa-\gamma\;,\quad
  \lambda_\mathrm{r}\equiv1-\kappa+\gamma
\label{eqn:eigenvalues}
\end{equation}
describe the distortion of an infinitesimal source in the tangential
and radial directions relative to the lens' centre and are thus called
\emph{tangential} and \emph{radial} eigenvalues. At points in the lens
plane where at least one of these two eigenvalues vanishes, the lens
mapping becomes singular, yielding formally infinite magnifications
for point sources. The set of all such points forms closed curves on
the lens plane, the \emph{critical curves}, whose images in the source
plane are the \emph{caustic curves} or \emph{caustics}.

The central quantity describing the statistics of strongly distorted
images with properties $p$ is the lensing cross section $\sigma_p$,
defined as the area of the region on the source plane where a source
with given characteristics has to lie in order to produce at least one
image with properties $p$. The rest of this paper deals with images
having length-to-width ratios exceeding a fixed threshold $d$. The
respective cross section will be denoted as $\sigma_d$.

This lensing cross section quantifies the probability for a set of
sources to be lensed by a given matter distribution into pronounced
arc-like images. If we want to know how many arcs are expected to form
on the entire sky (that is what we observe or can extrapolate from
observations), we must sum the cross sections of all individual
suitable lensing masses between us and the source plane, and then
allow for variations of the source redshift. The lensing optical depth
for sources at redshift $z_\mathrm{s}$ is
\begin{equation}
  \tau_d(z_\mathrm{s})=\frac{1}{4\pi D_\mathrm{s}^2}\,
  \int_0^{z_\mathrm{s}}\int_0^\infty\,
  N(M,z)\,\sigma_d(M,z)\,\d M\d z\;,
\label{eqn:depth}
\end{equation}
where $N(M,z)dz$ is the total number of masses $M$ contained in the
shell between redshifts $z$ and $z+\d z$. Denoting the differential
mass function by $n(M,z)$
\citep{PR74.1,BO91.1,LA93.1,LA94.1,SH99.1,SH02.1} and the cosmic
volume enclosed by a sphere of ``radius'' $z$ by $V(z)$, we can write
\begin{equation}
  N(M,z)=\left|\frac{\d V(z)}{\d z}\right|\,n(M,z)\;.
\end{equation}
If the number of sources with redshift between $z_\mathrm{s}$ and
$z_\mathrm{s}+\d z_\mathrm{s}$ is
$n_\mathrm{s}(z_\mathrm{s})\d z_\mathrm{s}$,
the total number of gravitational arcs is
\begin{equation}
  N_d=\int_0^\infty\,n_\mathrm{s}(z_\mathrm{s})\,
  \tau_d(z_\mathrm{s})\,\d z_\mathrm{s}\;.
\end{equation}
In particular, if all sources are at the same redshift $z_\mathrm{s}$,
the previous equation reduces to
\begin{equation}
  N_d=n_\mathrm{s}(z_\mathrm{s})\,\tau_d(z_\mathrm{s})\;.
\end{equation}
It is important for our discussion that the lensing cross section
$\sigma(M,z)$ of a single structure depends sensitively on the source
distribution and on the
internal properties of the structure itself,
in particular the slope of the inner density
profile, the lumpiness and the asymmetries of the matter distribution
\citep{BA95.1,BA98.2,ME03.1}. These peculiarities of galaxy clusters
are intimately linked with the formation and evolution histories of
the respective dark-matter haloes, which in turn depend on cosmology
(especially on the matter density parameter and the cosmological
constant or dark energy). Moreover, the spatial volume per unit
redshift, the source population and the cluster mass
function also depend on cosmology
\citep{LA93.1}.

Given the importance of asymmetries in cluster mass distributions it
is not promising to search an analytic expression for the function
$\sigma(M,z)$, unless we consider highly symmetric (and thus
unrealistic, see \citealt{ME03.1}) lens models. The alternative that
we shall discuss in the next subsections is to use numerical or
semi-analytic algorithms for calculating the lensing cross section of
a set of model clusters that appropriately sample the mass and
redshift range under consideration.

\section{Lensing Cross Sections}

So far, the most reliable method for calculating cross sections for
long and thin arcs has been using fully numerical ray-tracing
simulations. If performed adequately, such simulations return
realistic cross sections, but with the disadvantage of being very
expensive in terms of computational time. In view of cosmological
applications, the mass and redshift ranges to be covered are large, in
particular because the temporal sampling needs to be dense for
properly resolving cluster merger events. The number of cross sections
to be calculated can thus be very large. A reliable alternative to the
costly ray-tracing simulations is therefore needed. We develop here a
semi-analytic method which fairly reproduces fully numerical lensing
cross sections, but at a computational cost which is lowered by
factors of $\sim30$. We believe this method to provide an elegant
alternative to ray-tracing simulations for many applications of
lensing statistics.

In the following subsections, we shall first describe the fully
numerical method for reference, and then the semi-analytic method.

\subsection{Ray-Tracing Simulations}

Given the lensing mass distribution
(which can be given as either an analytic
density profile or a simulated density map) and the statistical properties of
the source sample, strong-lensing cross sections are commonly
estimated using fully numerical ray-tracing simulations. The method we
use here was first described by \cite{MI93.1} and subsequently further
developed and adapted to asymmetric, numerical lens models by
\cite{BA94.1}. It has been widely used by \cite{ME00.1,ME03.1} and
with several modifications by \cite{PU05.1}. We only address its
principal features here, referring the interested reader to the cited
papers and the references given therein for further detail.

Briefly, a bundle of $n\times n$ light rays ($n=2048$ here) with an
opening angle $\beta$ is sent from the observer through the lens. The
opening angle depends on the lens' properties and the distances
involved, and must be large enough to encompass the entire region on
the lens plane where strong lensing events may occur. The deflection
angle is calculated from the surface-mass distribution at all points
where light rays intersect the lens plane, thus allowing each ray to
be propagated back to the source plane by means of the lens equation
(\ref{eqn:lens}).

A set of sources is then placed on a regular and adaptive grid on the
source plane. The sources are modelled as ellipses whose orientation
angles and axis ratios (minor to major) are randomly drawn from the
intervals $[0,\pi]$ and $[0.5,1]$, respectively, with the prescription
each source has the area of a circle with $1$ arc second diameter. The
source-grid resolution is iteratively
increased near the caustics, i.e.~where the
magnification on the source plane is highest and undergoes rapid
variations.
This artificial increase in the probability of strong lensing events,
must be corrected when calculating cross sections.
We do so by assigning a statistical weight to each source which is proportional
to the area of the grid cell which it represents.

The images of each source are found by identifying all rays of the
bundle falling into the source.  Simple geometrical shapes (ellipses,
rectangles, and rings) are then fit to all images, and their
characteristics (length, width, curvature radius, etc.) are
determined. When an image has the property we are interested in
(i.e.~a length-to-width ratio equal to or greater than $d$), we
increment the cross section by the pixel area of the source grid,
times its statistical weight.

\subsection{Semi-Analytic Method}

\subsubsection{Point Sources}

We aim at determining cross sections for the formation of
gravitational arcs with length-to-width ratio exceeding some fixed
threshold $d$. Initially assuming infinitesimal (or point-like)
sources, the discussion in Sect.~2 implies that such arcs will form
where the ratio
\begin{equation}
  R(\vec x)\equiv\max\left[\left|
    \frac{\lambda_\mathrm{r}(\vec x)}{\lambda_\mathrm{t}(\vec x)}
    \right|,\left|
    \frac{\lambda_\mathrm{t}(\vec x)}{\lambda_\mathrm{r}(\vec x)}
    \right|\right]
\label{eq:ratio}
\end{equation}
between the eigenvalues of the lens mapping satisfies
\begin{equation}
  R(\vec x)\ge d\;.
\label{eqn:shell}
\end{equation}
We denote this region by $B_l=B_l(d)$. By means of the lens equation,
$B_l$ can be mapped onto an equivalent region $B_s=B_s(d)$ on the
source plane, whose area is by definition (see again Sect.~2) the
cross section $\sigma_d$ we are searching for. Thus,
\begin{equation}
  \sigma_d=\int_{B_s}\d^2\eta=\eta_0^2\int_{B_s}\d^2y\;.
\end{equation}
The lens equation maps the infinitesimal area element on the lens
plane to that on the source plane by means of the Jacobian determinant
$\det\mathcal{A}(\vec x)$, thus
\begin{equation}
  \sigma_d=\eta_0^2\int_{B_l}|\det\mathcal{A}(\vec x)|\d^2x=
  \eta_0^2\int_{B_l}\frac{\d^2x}{|\mu(\vec x)|}\;,
\label{eqn:sigma}
\end{equation}
where Eq.~(\ref{eqn:magnification}) was used.

\subsubsection{Extended Circular Sources}

Although this line of reasoning is exact, it fails in reproducing
simulated cross sections, for it does not account for real sources
(and also the sources used in ray-tracing simulations) not being
point-like. Extended sources are much more likely to produce strongly
distorted images than point-like sources, and their imaging properties
will only be approximated by the eigenvalue ratio $R(\vec x)$. This
implies that the integral in (\ref{eqn:sigma}) has to be extended
beyond the region $B_l$, because an extended source can produce a long
and thin arc even if it is centred where the eigenvalue ratio is less
than $d$.

Introducing extended sources into the framework just outlined, we wish
to convolve the eigenvalue ratio $R(\vec x)$ on the lens plane with a
suitable function to be determined which significantly differs from
zero only on the image of an extended source. For consistency with
ray-tracing simulations, we assume circular sources with radius
$\zeta=0.5''$ for now. Effects of source ellipticities will be
included later.

The obvious problem with this approach is that the properties of the
image vary wildly across the lens plane, thus the convolution should
use a function which rapidly changes shape and extent across the lens
plane. We can avoid this problem by transferring the calculations to
the source plane. We first introduce the eigenvalue ratio on the
source plane. Since multiple points $\vec x_i$ on the lens plane may
be mapped onto a single point $\vec y$ on the source plane, we define
it as $\bar R(\vec y)\equiv\max\{R[\vec y(\vec x_i)]\}$. This ratio
$\bar R(\vec y)$ is then convolved with a function $g(\vec y)$ which
differs significantly from zero only on the circular area covered by a
source, which is the same everywhere on the source plane. Let the
convolution in the source plane be $\bar h(\vec y)\equiv\bar R(\vec
y)*g(\vec y)$, then we put $h(\vec x)=\bar h[\vec y(\vec x)]$ to
obtain the convolved function on the lens plane, as desired. Again, we
need to take into account that single points on the source plane may
be mapped on multiple points in the lens plane. Assigning $R(\vec x)$
to $\bar R[\vec y(\vec x)]$ now sets the convolved length-to-width
ratio equal on all image points, which is not exact, but a tolerable
error. The problem is now reduced to carrying out the convolution
$\bar h(\vec y)$, which is discussed in the Appendix. As described
there, we speed up the convolution by approximating it by a simple
multiplication after essentially assuming that the eigenvalues of the
lens mapping and their ratio do not change significantly across
a single source.

Finally, what form should be chosen for the function $g(\vec y)$? A
two-dimensional Gaussian of width $\zeta$ may appear
intuitive, but the ray-tracing simulations we use for reference do not
adopt a surface-brightness profile for sources because they simply
bundle all rays falling into the ellipse representing a source. Thus,
a choice for $g$ consistent with the ray-tracing simulations is a step
function with width $\zeta$,
\begin{equation}
  g(\vec y)=\left\{\begin{array}{l}
  (\pi\zeta^2)^{-1}\quad
  \mbox{where}\quad\vec y^T\mathcal{B}\vec y\le1\\
  0\quad\mbox{else}\\
  \end{array}\right.\;,
\label{eqn:step}
\end{equation}
where $\mathcal{B}$ is a matrix describing the shape of the
source. Since we are only considering circular sources of radius
$\zeta$, $\mathcal{B}$ has the form
$\mathcal{B}=\mathcal{I}/\zeta^2$, where $\mathcal{I}$ is the
unit matrix. The factor $(\pi\zeta^2)^{-1}$ in (\ref{eqn:step})
normalises $g(\vec y)$ to unity. Substituting $h(x)$ for $R(x)$, we
can again apply Eq.~(\ref{eqn:shell}) to obtain a new region $B_l$
that will now of course be larger than before.

\subsubsection{Extended Elliptical Sources}

We finally have to account for elliptical rather than circular
sources.  It is quite obvious and was shown by many authors
(e.g.~\citealt{BA95.1,KE01.2}) that elliptical sources are more likely
to produce strongly distorted images.

We adopt the simple and elegant formalism by \cite{KE01.2}. He showed
that an elliptical source at position $\vec y(\vec x)$ with axis ratio
$q_s=a/b=q_s[\vec y(\vec x)]$ and orientation angle
$\theta=\theta[\vec y(\vec x)]$ is imaged as an allipse
with an axis ratio of
\begin{equation}
  q_\mathrm{obs}=\sqrt{\frac{T+\sqrt{T^2-4D}}{T-\sqrt{T^2-4D}}}
\end{equation}
where $T$ and $D$ are the trace and the determinant of the matrix that
describes the image ellipse. They can be expressed as functions of
the intrinsic source properties and of the (convolved) lensing distortion:
\begin{equation}
  T=h^2+q_s^2+(h^2-1)(q_s^2-1)\,\cos^2\theta\;,\quad
  D=h^2q_s^2\;.
\end{equation}
Again, $h(x)$ can be replaced by $q_\mathrm{obs}$ to obtain a
modification of the region $B_l$, again larger than before, whose area
is now a close approximation to the cross section we seek to
determine, accounting for extended and elliptical sources.

\subsubsection{Comparison to Ray-Tracing}

\begin{figure*}[ht]
\begin{center}
  \includegraphics[width=0.3\hsize]{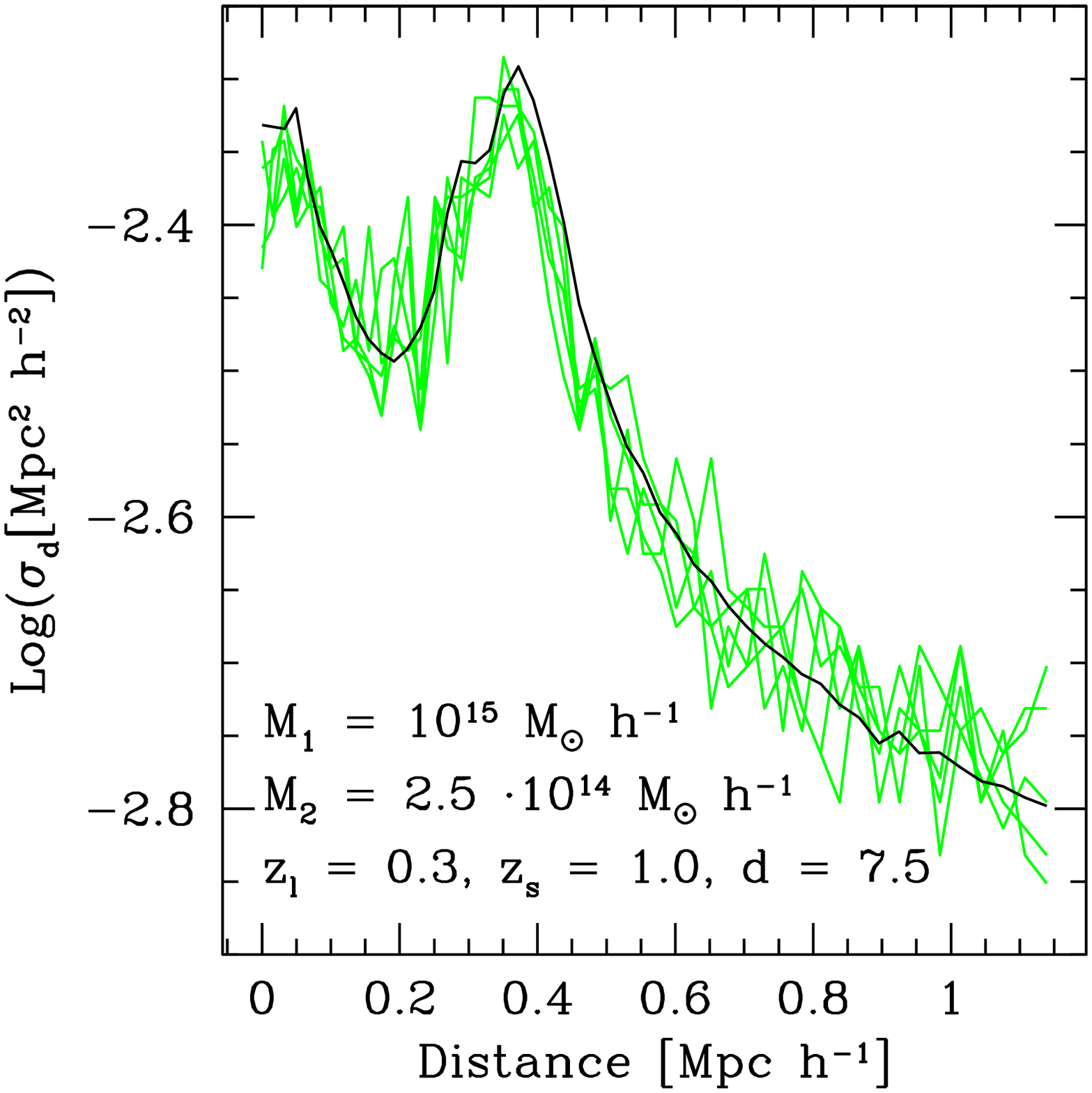}
  \includegraphics[width=0.3\hsize]{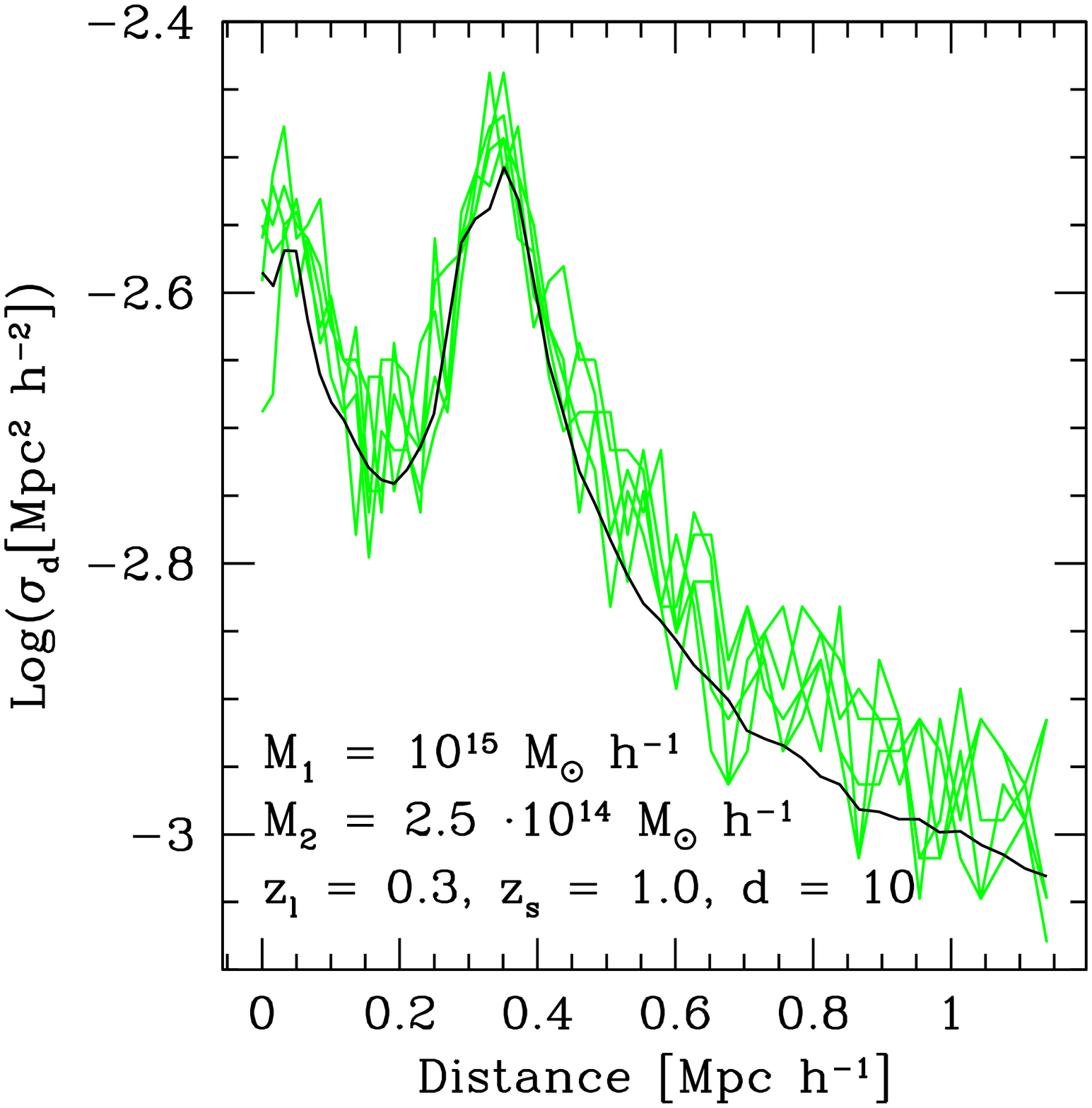}\\
  \includegraphics[width=0.3\hsize]{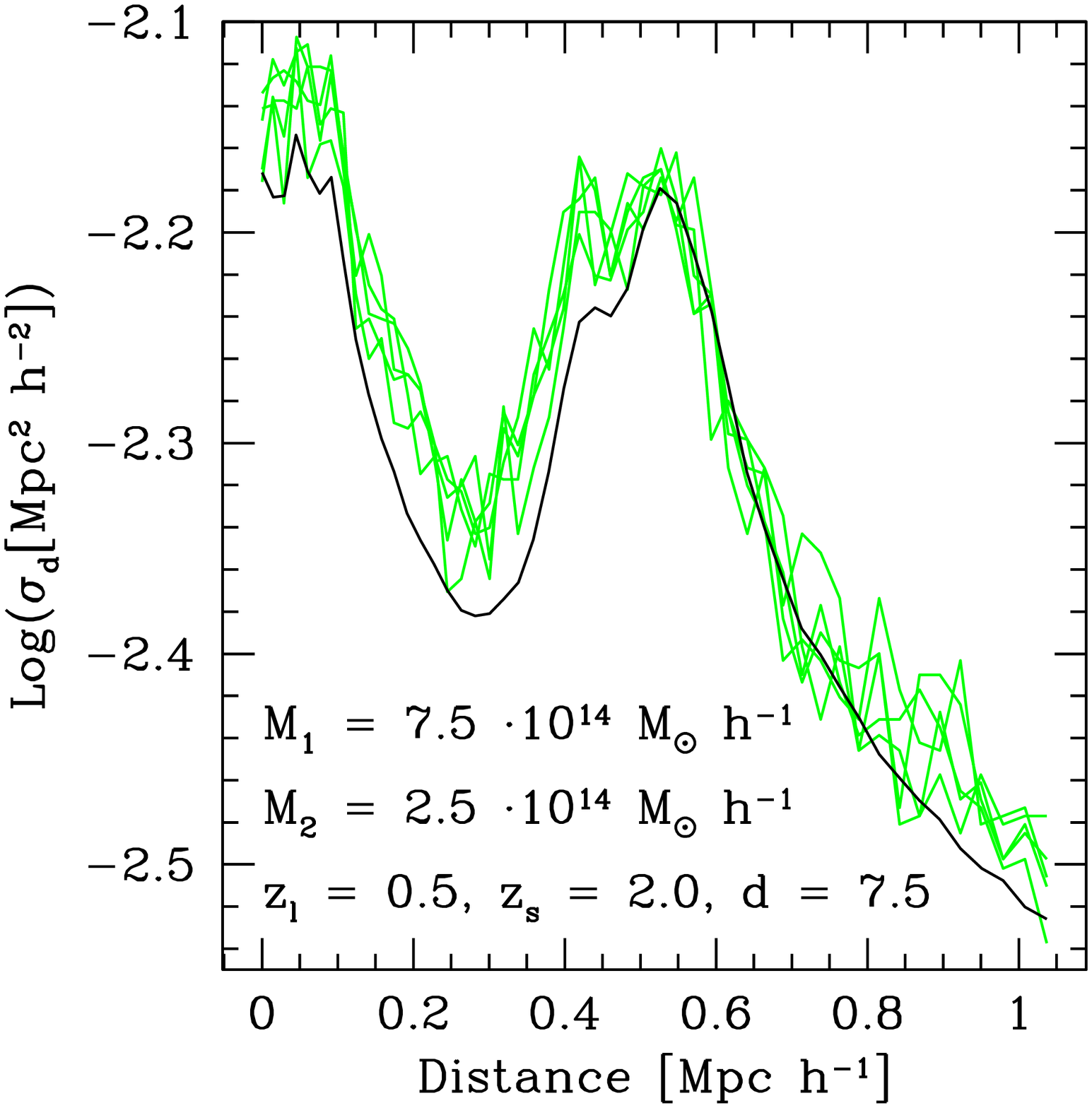}
  \includegraphics[width=0.3\hsize]{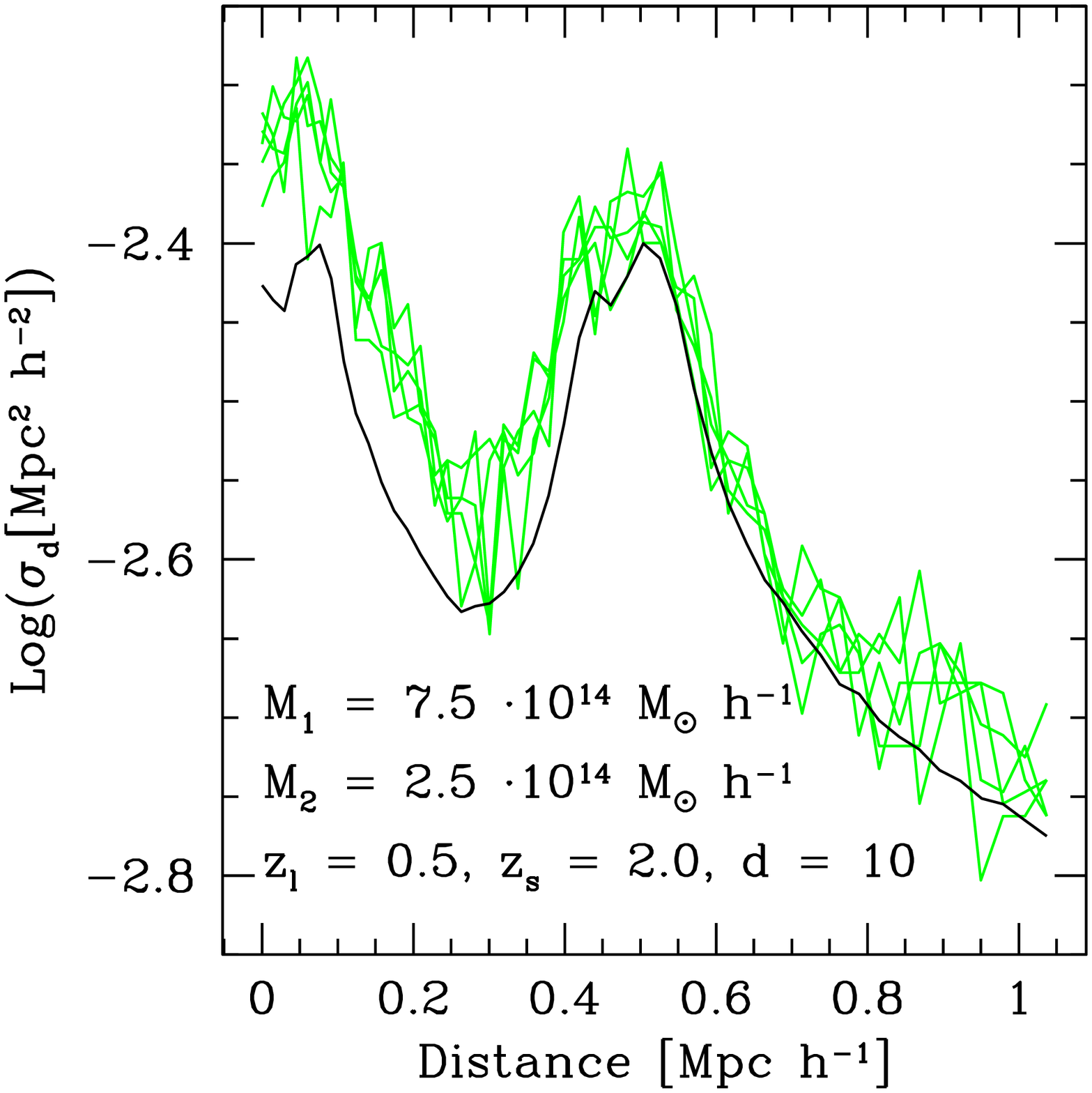}
\end{center}
\caption{Lensing cross sections for long and thin arcs with
  length-to-width ratio exceeding $d$ as a function of the distance
  between the centres of two dark matter haloes, for two (lens and
  source) redshifts. The haloes are modelled as NFW density profiles
  with elliptically distorted lensing potential with ellipticity
  $e=0.3$ (see Sect.~4).
  Green lines show the results of ray-tracing simulations
  for five different realisations of the source distribution. Black
  lines are the results obtained with our semi-analytic method.}
\label{fig:crsec}
\end{figure*}

As a final step, we test the accuracy of our calculations and
approximations, especially those concerning the replacement of the
convolution by a multiplication (cf.~the Appendix), by comparing
lensing cross sections obtained with the semi-analytic method to their
fully numerical counterparts obtained by ray-tracing. Since the main
purpose of this work is to estimate the effects of cluster mergers, we
compare the cross sections as they evolve while two dark matter haloes
merge. The halos are modelled as NFW profiles (see
\citealt{NA95.3,NA96.1,NA97.1,CO96.1}) whose lensing potential is
elliptically distorted to have ellipticity $e=0.3$ (see Sect.~4 for
detail). The results of the comparison are shown in
Fig.~\ref{fig:crsec} for various masses, redshifts and thresholds $d$.

Figure~\ref{fig:crsec} warrants one observation. The
ray-tracing simulation was repeated five times for each modelled merger
event, each time changing the seed for the random generation of source
ellipticities and orientations. As the figure shows,
this has quite a significant effect on the numerical cross sections,
causing a substantial scatter. The reason can be easily understood. When a
given source produces an arc, variations in the source's intrinsic ellipticity
and in the orientation with respect to the caustic structure can easily push
the length-to-width ratio of the image arc below or above the chosen threshold
$d$.
This causes the irregular trend seen in Figure~\ref{fig:crsec}. It is worth
noting that the same problem in principle also affects the semi-analytic
results. In this case, however, ellipticity and orientation are assigned to
each ray traced back through the lens plane simply to identify the
the extended region over which we integrate. Since the number of rays is
much larger than the number of individual sources used in the fully numerical
algorithm, the random scatter in the semi-analytic results is
very small. In fact, the fluctuations are of the order of the width
of the black curve and thus omitted.

Moreover, it must be noted that the results given by the ray-tracing code
we use might differ slightly from other codes using different resolution,
different image finding algorithms and different ways to fit the image shapes.

Analysing Fig.~\ref{fig:crsec}, we can draw two interesting
conclusions. The first is that, reassuringly, the numerical and the
semi-analytic cross section agree excellently. This means that the
approximations we made are substantially correct. There is, however, a
small discrepancy for higher-redshift sources, shown in the lower
panels. We believe that this discrepancy is due to the fact that the
scale over which the lensing properties on the source plane change
significantly is in this case comparable to the source dimension, thus
one of the approximations we made in the Appendix fails, namely that
the function $R[\vec y(\vec x)]$ does not vary much across a
source. Nonetheless, the discrepancy is nowhere larger than 20\%,
which is more than acceptable for our purposes, in particular in view
of the considerable scatter in the ray-tracing results.

The second observation is that the behaviour of the cross sections
(both numerical and semi-analytic) as a function of the distance
between the haloes closely reflects that found by \cite{TO04.1}. These
authors studied the effect of a single merger process on the
strong-lensing efficiency of a numerically simulated dark matter halo,
using ray-tracing as described in Sect.~3. They found that, while the
substructure is swallowed by the main halo, there is a first peak in
the lensing cross section when the increasing shear field between the
lumps causes the critical lines to merge, a local minimum while cusps
disappear in the caustic branches, and another peak caused by the
increased convergence when the two density profiles overlap. All these
features are recovered in the panels of the Fig.~\ref{fig:crsec}.

This level of agreement between ray-tracing and semi-analytic cross
sections show that the semi-analytic method developed above for
cross-section calculations is essentially correct and constitutes a
valid and useful, $\sim30$ times faster alternative to the costly
ray-tracing simulations.

\section{Lensing Optical Depth}

Given reliable lensing cross sections, lensing optical depths need to
be determined following Eq.~(\ref{eqn:depth}). For that purpose, we
use the merger trees for a set of 46 numerically simulated dark-matter
haloes, whose main properties are described in the next subsection.
The merger tree of a dark-matter halo provides two types of
information. First, we know how the mass of each halo evolves with
time or redshift. Second, we know at which redshift merger events
happen with substructures of known mass.  Regarding this, it is worth
recalling that the evolution of a dark-matter halo is characterised by
the continuous accretion of infalling external material. We account for
all mergers in which the main halo accretes a sub-haloes with at least
5\% of the main halo's mass. Nonetheless, it should be kept in mind
that halos continuously accrete matter apart from merger events.

\subsection{Halo Model}

We base our study on the merger trees on a sample of 46 numerically
simulated dark-matter haloes. These haloes were re-simulations at
higher resolution of a large-scale cosmological simulation. The
cosmological model used was a ``concordance'' $\Lambda$CDM model,
having a cosmological constant of $\Omega_{\Lambda,0}=0.7$, a (dark)
matter density of $\Omega_{\mathrm{m},0}=0.3$ and a Hubble constant of
$h=0.7$. The power spectrum of the primordial density-fluctuation
field is scale invariant (i.e.~with a spectral index of $n=1$), and
the \emph{rms} linear density fluctuations on a comoving scale of
$8\,h^{-1}\,\mathrm{Mpc}$ is $\sigma_8=0.9$, which is the typical
value required to match the local abundance of massive galaxy clusters
\citep{WH93.1,EK96.1}. The mass of dark-matter particles is
$m=1.3\times10^9\,h^{-1}\,M_\odot$. In Tab.~\ref{tab:1} we list the
present masses and virial radii for the three most massive and the
tree least massive haloes in the sample.

\begin{table}
\caption{Present-day masses and virial radii for the three most
  massive and the three least massive haloes in our sample.}
\label{tab:1}
\begin{center}
\begin{tabular}{|l|r|r|}
\hline
Halo Id. & Virial Mass & Virial Radius\\
 & [$h^{-1}\,M_\odot$] & [$h^{-1}\,\mathrm{Mpc}$]\\
\hline
\hline
g8-a & 2.289 $10^{15}$ & 2.146 \\
g1-a & 1.530 $10^{15}$ & 1.876 \\
g72-a & 1.374 $10^{15}$ & 1.810 \\
\hline
g696-y & 5.219 $10^{13}$ & 0.608\\
g696-z & 5.171 $10^{13}$ & 0.607\\
g696-\# & 5.060 $10^{13}$ & 0.602\\
\hline
\end{tabular}
\end{center}
\end{table}

The present masses of the halo models vary between about
$5\times10^{13}\,h^{-1}\,M_\odot$ (just exceeding the mass of a
massive cD galaxy) and more than $2\times10^{15}\,h^{-1}\,M_\odot$,
which is characteristic of a rich galaxy cluster.

We note in passing that, owing to the dependence of the
cluster-evolution history on cosmology, the typical redshift of
structure formation in the cosmological model used here is higher than
in an Einstein-de Sitter universe, and lower than in an open
low-density universe. Models of dark energy alternative to the
cosmological constant typically shift this redshift towards higher
redshifts, thus changing the contribution of cluster mergers to arc
statistics. We are planning a future study to address this interesting
property.

For describing specifically each individual merger process and its
effects on gravitational-arc statistics, we model each dark-matter
halo as NFW spheres. Their density profile is
\begin{equation}\label{eqn:nfw}
  \rho(r)=\frac{\rho_\mathrm{s}}
               {r/r_\mathrm{s}(1+r/r_\mathrm{s})^2}\;,
\end{equation}
which fits the density profiles of numerically simulated dark-matter
haloes and fairly describes dark structures over a wide range of
masses and in different cosmologies.  The scale density
$\rho_\mathrm{s}$ and the scale radius $r_\mathrm{s}$ are linked by
the halo concentration, $c=r_{200}/r_\mathrm{s}$, where $r_{200}$ is
the radius enclosing a mean density of 200 times the closure density
of the Universe. As a general trend, the concentration reflects the
mean density of the Universe at the collapse redshift of the halo,
thus it is higher for lower masses. Several ways to quantitatively
relate the concentration to the halo mass have been proposed. We
follow the prescription of \cite{EK01.1}. The analytic fit
(\ref{eqn:nfw}) allows direct calculations of the main lensing
properties. In particular, the projected density profile for a NFW
lens is
\begin{equation}
  \Sigma(x)=2\rho_\mathrm{s}r_\mathrm{s}\frac{f(x)}{x^2 - 1}\;,
\end{equation}
and its deflection angle is
\begin{equation}
  \alpha(x)=4\rho_\mathrm{s}r_\mathrm{s}\frac{g(x)}{x}\;,
\label{eqn:deflection}
\end{equation}
where the functions $f(x)$ and $g(x)$ are
\begin{eqnarray}
  f(x)&=&\left\{\begin{array}{l}
    1-\frac{2}{\sqrt{x^2-1}}\arctan{\sqrt{\frac{x-1}{x+1}}}
    \quad\hbox{if}\quad x>1\\
    0\quad\hbox{if}\quad x=1\\
    1-\frac{2}{\sqrt{1-x^2}}\mbox{arctanh}{\sqrt{\frac{1-x}{1+x}}}
    \quad\mbox{else}\\
\end{array}\right.\;,\nonumber\\
  g(x)&=&\ln\frac{x}{2}+\left\{\begin{array}{l}
  \frac{2}{\sqrt{x^2-1}}\arctan{\sqrt{\frac{x-1}{x+1}}}
  \quad\mbox{if}\quad x>1\\
  1\quad\mbox{if}\quad x=1\\
  \frac{2}{\sqrt{1-x^2}}\mbox{arctanh}{\sqrt{\frac{1-x}{1+x}}}
  \quad\mbox{else}\\
\end{array}\right.
\end{eqnarray}
\citep{BA96.1}. Here and in the remainder of this sub-section, $x$ is
the dimensionless distance from the lens' centre. We additionally
adopt elliptically distorted lensing potentials. Elliptical potentials
are not equivalent to elliptical density distributions, but allow for
much more straightforward calculations \citep{SC92.1}, and the results
are equally valid for our purposes \citep{GO02.1,ME03.1}. In their
work, \cite{ME03.1} also estimated the value of the ellipticity $e$
for the lensing potential that produces the best fit to the
deflection-angle maps of simulated haloes. Following their result, we
adopt $e=0.3$ throughout.

\subsection{Modelling Halo Mergers}

Once the redshift is fixed, the halo we consider may be isolated or in
interaction with a substructure. In the first case the deflection
angle can be obtained directly by means of equation
(\ref{eqn:deflection}). In the second, we can sum the deflection
angles of the two structures at each point on the lens plane, owing to
the linearity of the problem \citep{SC92.1,TO04.1}. In both cases, the
eigenvalues of the local mapping follow from the deflection-angle maps
by differentiation, and can be applied in the semi-analytic method for
calculating cross sections as described in Sect.~3.

Individual merger events are modelled as follows: The main halo and
the substructure start at a mutual distance equal to the sum of their
virial radii. Then the centres approach at a constant speed,
calculated by the ratio of the initial distance to the typical time
scale of a merger event, i.e.~0.9~Gyr \citep{TO04.1,TO04.2}. The
process is assumed to be concluded when the two density profiles
overlap perfectly. The direction of approach, always assumed to be
perpendicular to the observer's line of sight, is parallel to the
directions of the major axes of the lensing potential, which are also
assumed to be parallel. This last assumption is well justified by
recent work \citep{LE05.1,HO05.2} pointing out, both with analytic
models and numerical analyses, that there is an intrinsic alignment
between a dark-matter halo and the surrounding haloes and sub-lumps
due to the tidal field of the major halo itself. We note that assuming
all mergers to proceed along directions perpendicular to the
line-of-sight slightly \emph{underestimates} the cumulative
contribution of cluster mergers to the lensing optical depth, because
if the merger direction is partially aligned with the line-of-sight,
the time spent by the system in configurations leading to high cross
sections is longer (see again Fig.~\ref{fig:crsecbis}). Several tests
we made show however that this underestimation is typically of order
20\%, thus we neglect it here.

Once we know for each halo at every redshift the cross
section with and without the effect of mergers, we can compute lensing
optical depths. Since we sample the mass range discretely,
we cannot apply Eq.~(\ref{eqn:depth}) directly, but need to
approximate it as
\begin{equation}
  \tau_d(z_\mathrm{s})=\int_0^{z_\mathrm{s}}\left[
  \sum_{i=1}^{n-1}\bar\sigma_{d,i}(z)\int_{M_i}^{M_{i+1}}
  N(M,z)\d M\right]\frac{\d z}{4\pi D_\mathrm{s}^2}\;,
\label{eqn:numdepth}
\end{equation}
where the masses $M_i$ ($1\le i\ne n$) have to be sorted from the
smallest to the largest at each redshift step, and the quantity
$\bar\sigma_{d,i}(z)$ is defined as
\begin{equation}
  \bar\sigma_{d,i}(z)=\frac{1}{2}\left[
  \sigma_d(M_i,z)+\sigma_d(M_{i+1},z)
  \right]\;.
\end{equation}

This is essentially equivalent to assigning to all structures with mass
within $[M_i,M_{i+1}]$ the average cross section of the haloes
with masses $M_i$ and $M_{i+1}$, weighted with the number density of
massive structures within that interval.

\subsection{Mass Cut-Off}

Since a sufficient condition for strong lensing is satisfied if the
surface density of a lens exceeds the critical density somewhere, any
lens model with a cuspy density profile such as the NFW profile will
formally be a strong lens and thus produce critical curves and
caustics. However, the caustics of low-mass halos will be smaller than
the typical source galaxies. Averaging the local distortion due to a
low-mass lens across an extended source will then lead to a small or
negligible total distortion. This implies that the mass necessary for
halos to cause large arcs is bounded from below by the requirement
that the halo's caustic must be sufficiently larger than the available
sources.

This mass limit obviously depends on the lens and source redshifts due
to the geometrical sensitivity of lensing. Higher masses are required
at low and high redshifts for lensing effects comparable to lenses at
intermediate redshifts. In addition, the caustic structures can change
substantially during major halo mergers. As a sub-halo approaches the
main halo, the initially separated caustics of the two merging
components will increase and merge to form a larger caustic. Thus,
even though the total mass is unchanged, the mass limit for strong
lensing may decrease while a merger proceeds. Even halos which are
individually not massive enough for arc formation may be pushed above
the mass limit while they merge. In view of the exponentially dropping
cluster mass function, this is potentially a huge effect.

\begin{figure}[ht]
  \includegraphics[width=\hsize]{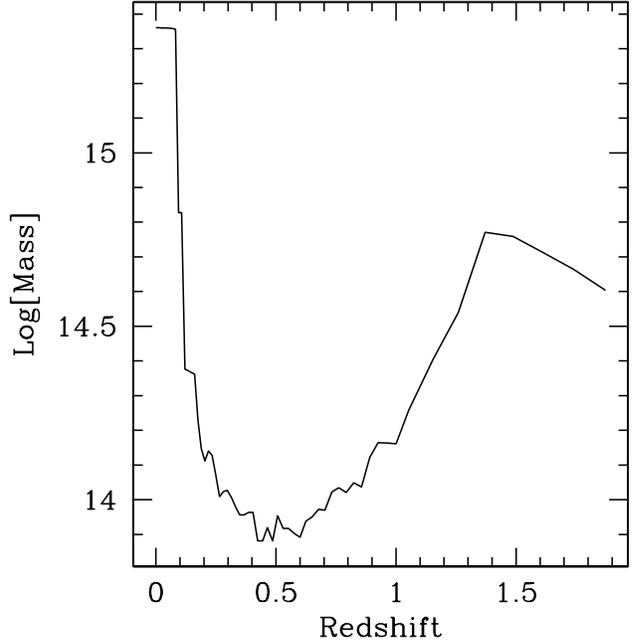}
\caption{Mass of the lowest-mass halo producing large arcs in the
  sample of 46 halos which we use here to compute the lensing optical
  depth. The source redshift is $z_\mathrm{s}=2$.
  The overall trend in the curve 
  reflects the geometrical
  lensing sensitivity, while the fluctuations and the depth of the
  minimum reflect that mergers can lift low-mass halos above the
  strong-lensing threshold which would otherwise not be capable of
  strong lensing.}
\label{fig:massmax}
\end{figure}

Thus, we have to monitor the extent of the caustics as we compute the
lensing optical depth of a halo sample, taking into account that the
mass limit may change rapidly as halos merge with sub-halos. The lowest
mass of a halo from our sample of 46 halos which still contributes to
strong lensing is shown in Fig.~\ref{fig:massmax} as a function of
lens redshift with fixed source redshift $z_\mathrm{s}=2$.

Following these prescriptions, we are able to calculate the
strong-lensing cross section of each model halo at every redshift step
of the simulation, first ignoring the effects of merger processes,
then accounting for them. Thus, we sort the masses from the smallest
to the largest at every redshift step and calculate the quantities
$\sigma_{d,i}(z)$. Finally, we calculate the lensing optical depth
with and without the effect of merger processes.

\section{Results}

\subsection{Cross Sections}

We calculate the optical depth for the formation of gravitational arcs
with length-to-width ratio grater than an arbitrary threshold $d$. We
used two choices, $d=7.5$ and $d=10$. 

\begin{figure*}[ht]
\begin{center}
  \includegraphics[width=0.225\hsize]{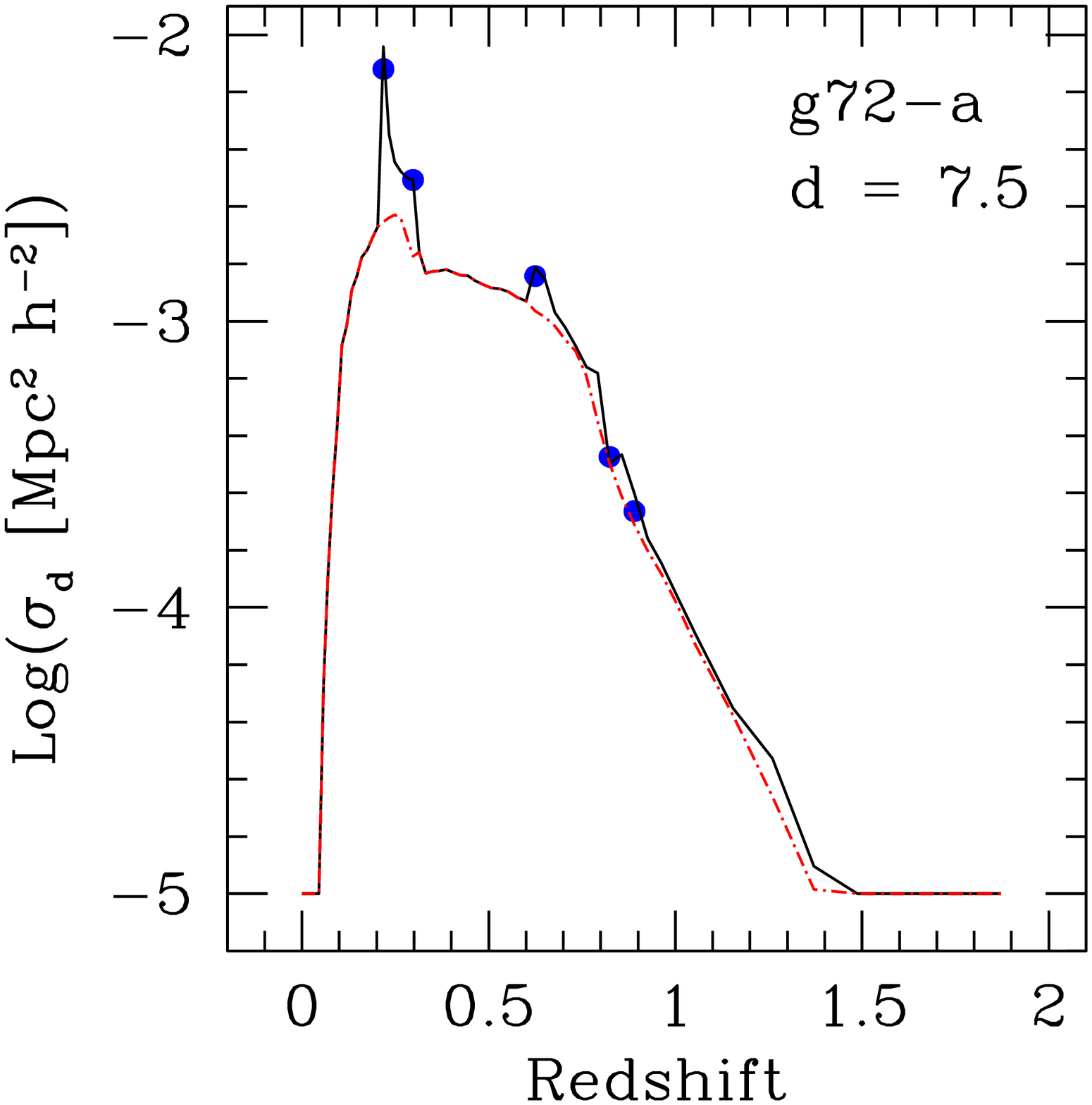}
  \includegraphics[width=0.225\hsize]{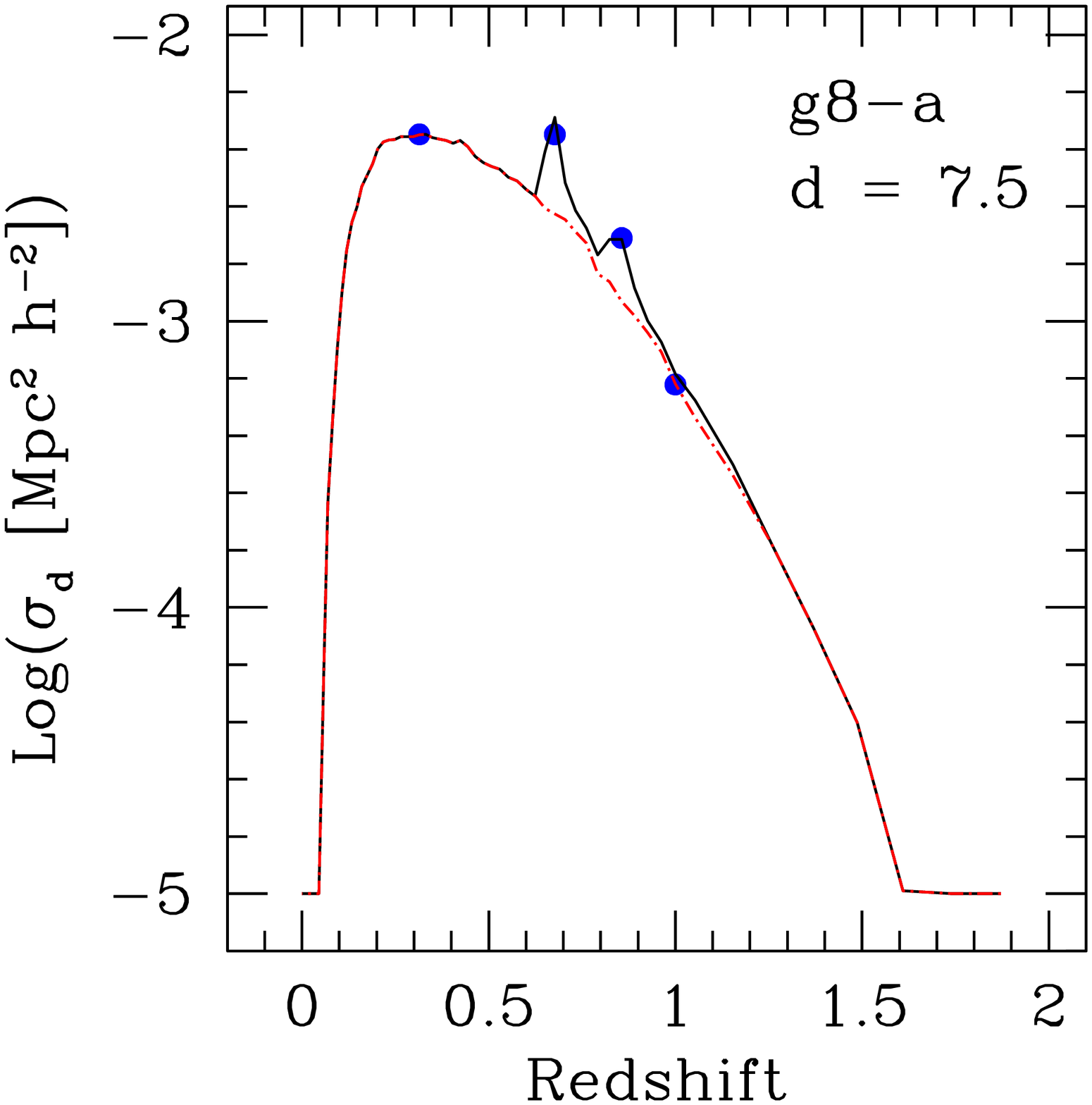}
  \includegraphics[width=0.225\hsize]{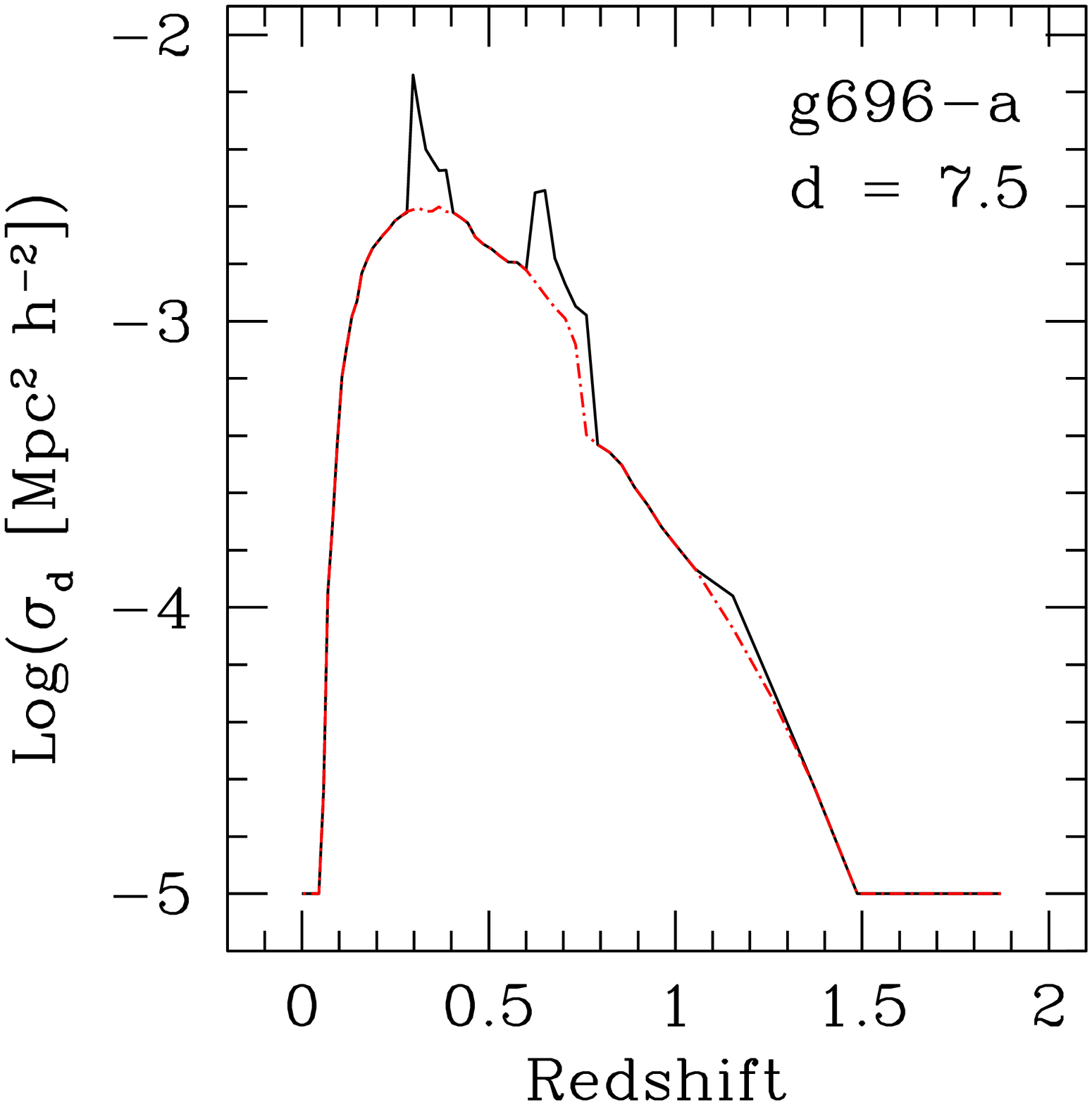}
  \includegraphics[width=0.225\hsize]{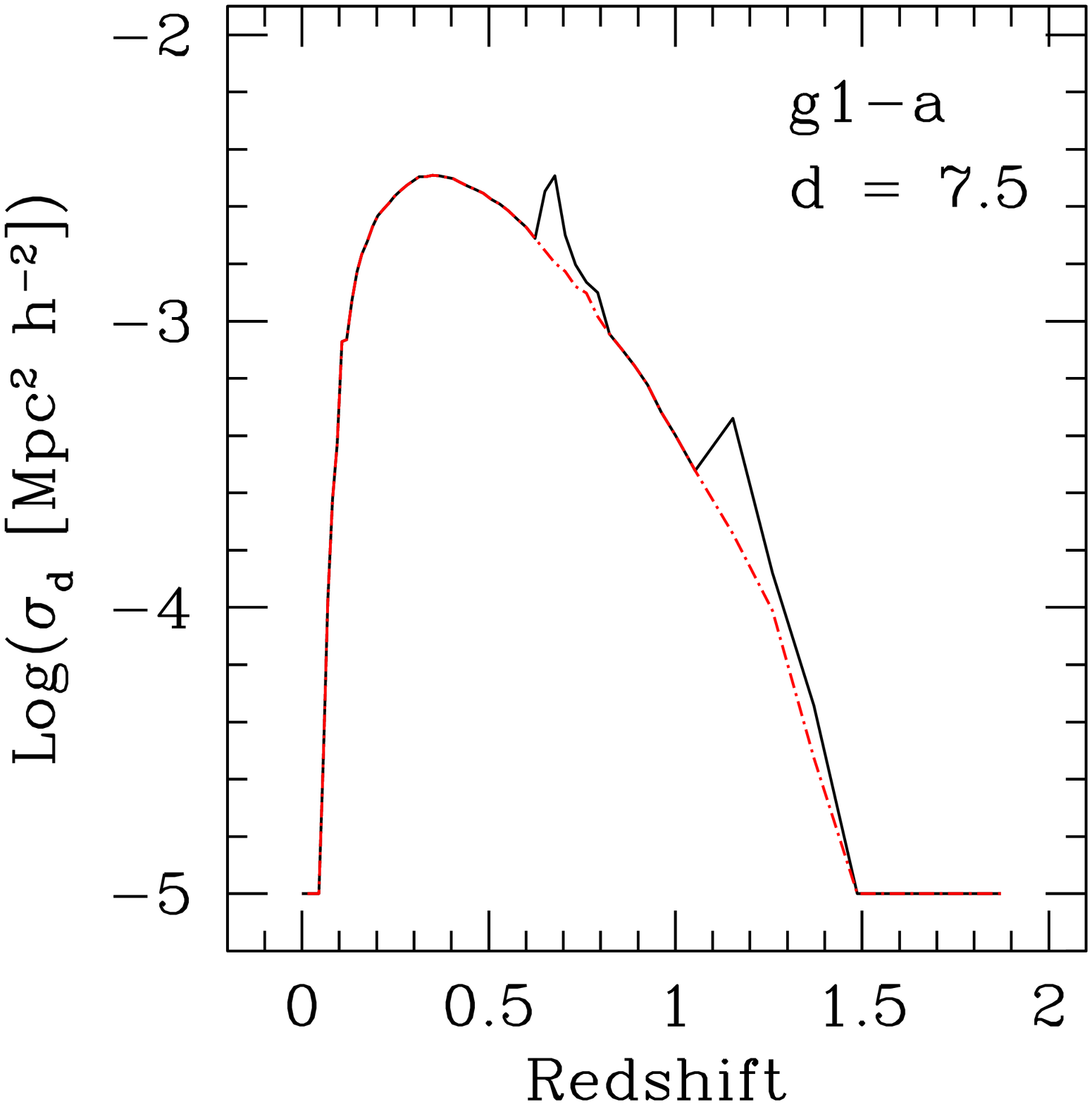}\\
  \includegraphics[width=0.225\hsize]{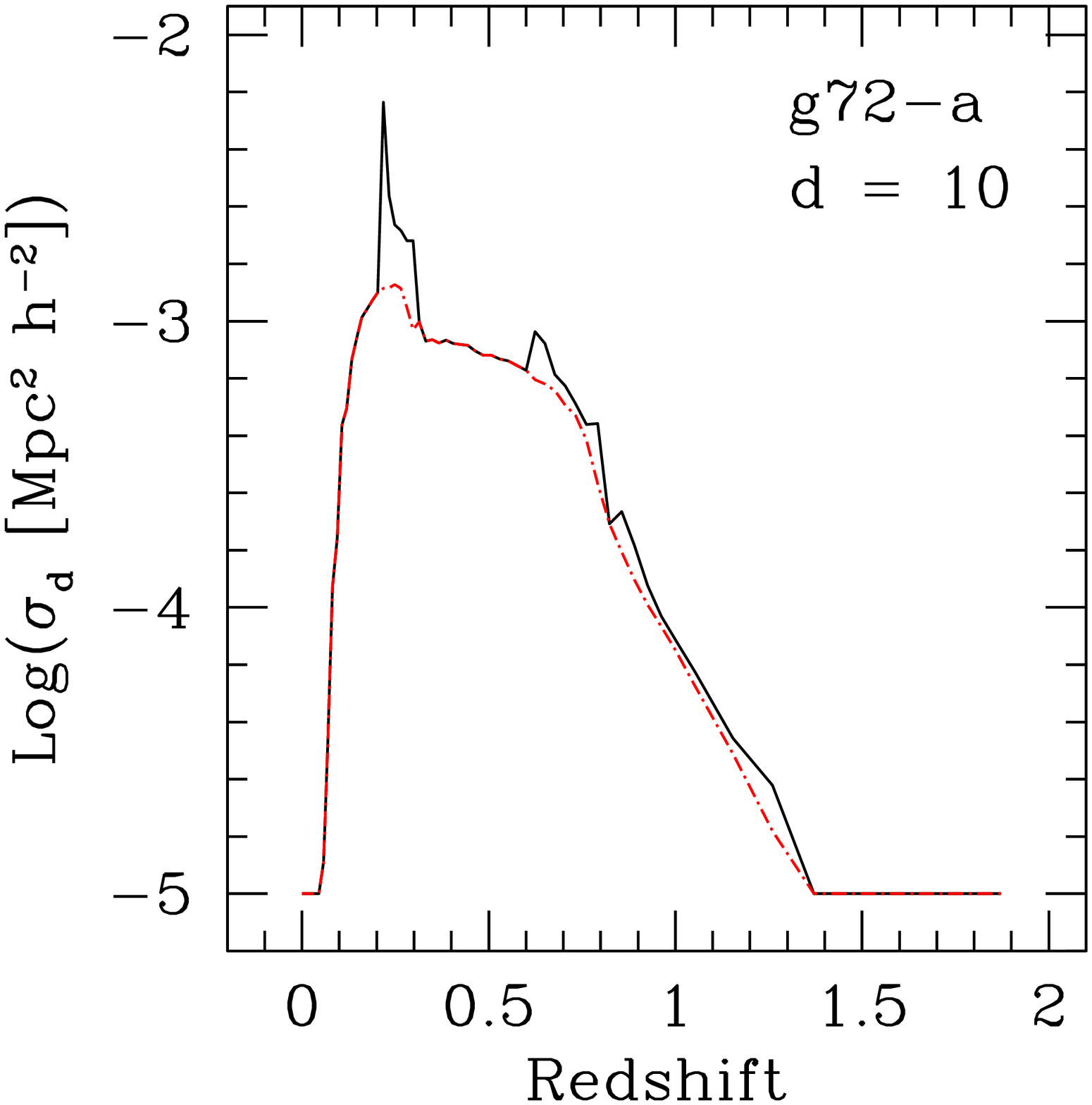}
  \includegraphics[width=0.225\hsize]{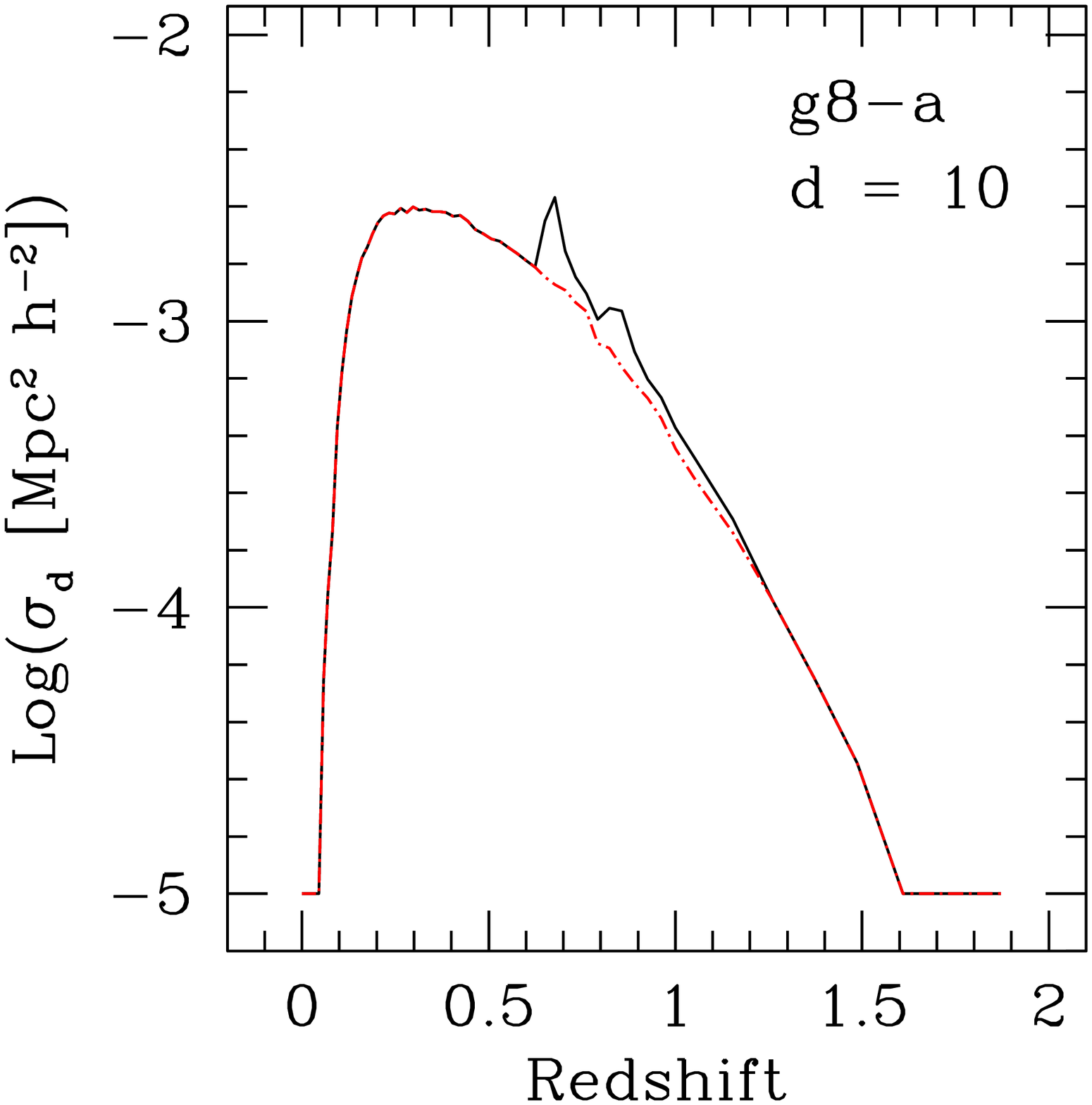}
  \includegraphics[width=0.225\hsize]{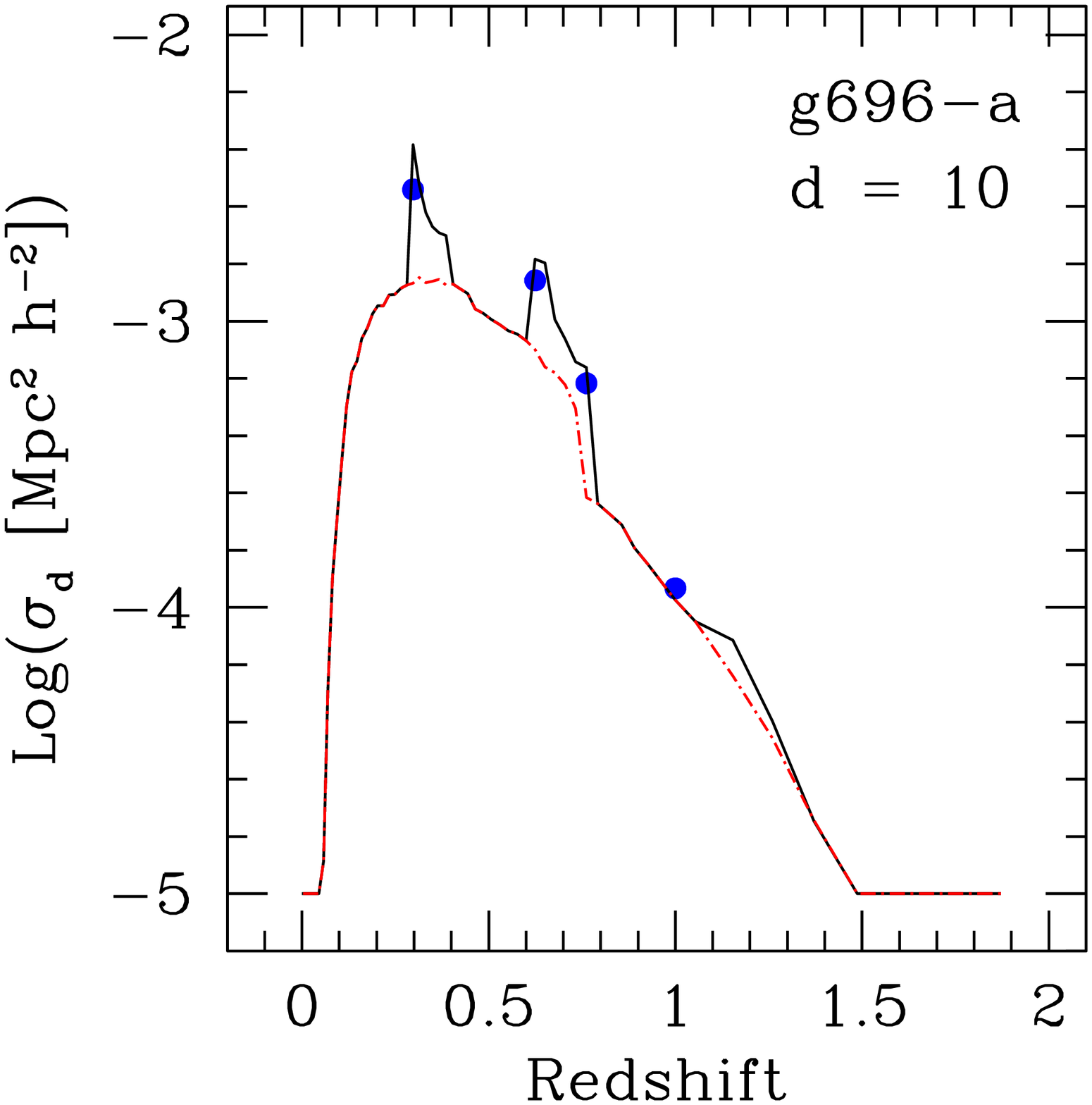}
  \includegraphics[width=0.225\hsize]{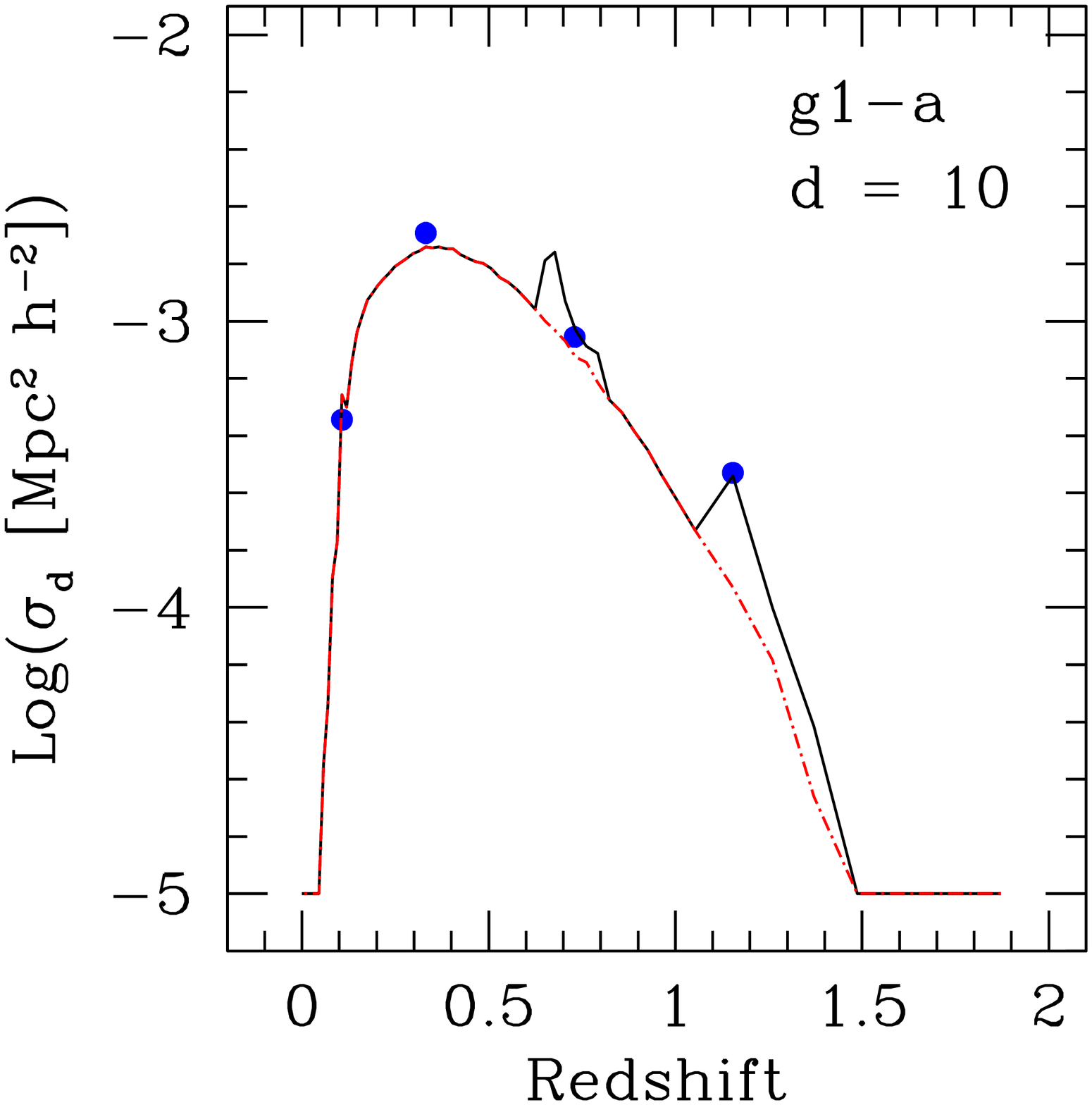}
\end{center}
\caption{Evolution of the lensing cross section for gravitational arcs
  with length-to-width ratio equal to or greater than $d=7.5$ (top
  panels) and $d=10$ (bottom panels) for four of the most massive
  haloes in the sample. Sources are at redshift $z_\mathrm{s} = 2$.
  Red-dashed lines show cross sections
  calculated without taking account of merger processes. Black lines
  show cross sections enhanced by cluster interactions. Filled blue
  dots are the counterparts of the black lines obtained from fully
  numerical ray-tracing simulations.}
\label{fig:crsecbis}
\end{figure*}

Before discussing the
calculation of the optical depth it is interesting and useful to study
the behaviour of lensing cross sections of individual haloes with
redshift.  In Fig.~\ref{fig:crsecbis} we show the cross sections for
four of the most massive haloes, both with $d=7.5$ (top panels) and
$d=10$ (bottom panels). The sources are put at a fixed redshift
$z_\mathrm{s}=2$. 
Apart from the obvious fact that cross
sections for arcs with higher length-to-width ratio are smaller than
those including shorter arcs, we see that all cross sections tend to
zero when the redshift approaches zero or the source redshift. This
``suppression'' due to the lensing efficiency is caused by the
geometry of the problem, and in particular by the fact that lenses
very close to the sources or the observer have arbitrarily large
critical density. Moreover we note that the increase in the cross
section due to merger events can exceed half an order of magnitude in
some cases. \cite{TO04.1} found an increase in the lensing cross
section up to an order of magnitude, but here, even if mergers happen
at redshifts with higher lensing efficiency ($z_\mathrm{l}\simeq0.2-0.5$), the
masses involved are lower, so we cannot reach such higher
increases. In particular, high-redshift mergers are quite inefficient
in boosting total lensing cross sections, both because of low involved
masses and proximity to the source plane.  To further test the
reliability of our semi-analytic calculations we show as filled blue
dots in Fig.~\ref{fig:crsecbis} the cross sections obtained from
ray-tracing simulations. The agreement is again reassuringly good.

\subsection{Optical Depths}

\begin{figure}
  \includegraphics[width=\hsize]{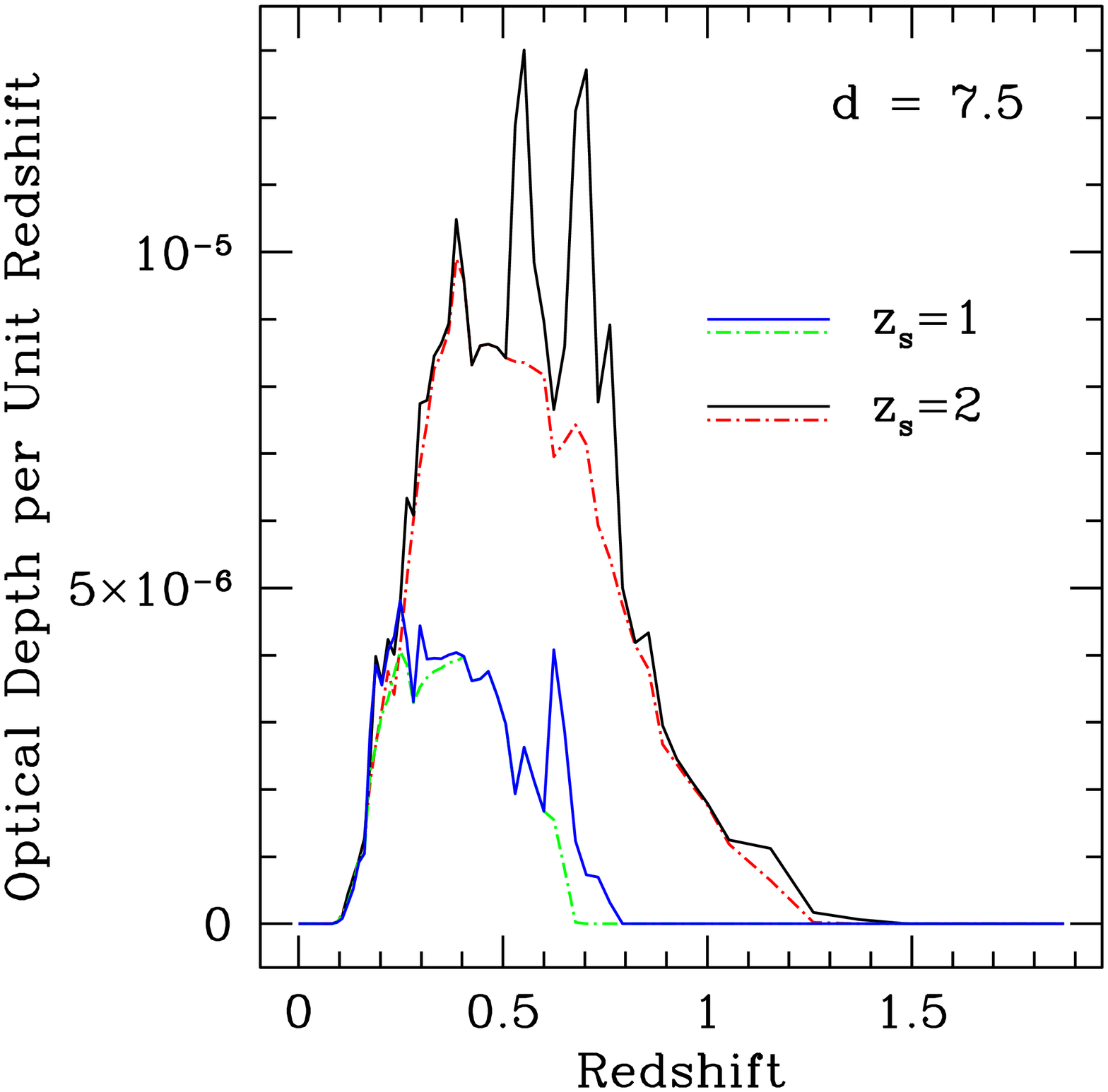}
  \includegraphics[width=\hsize]{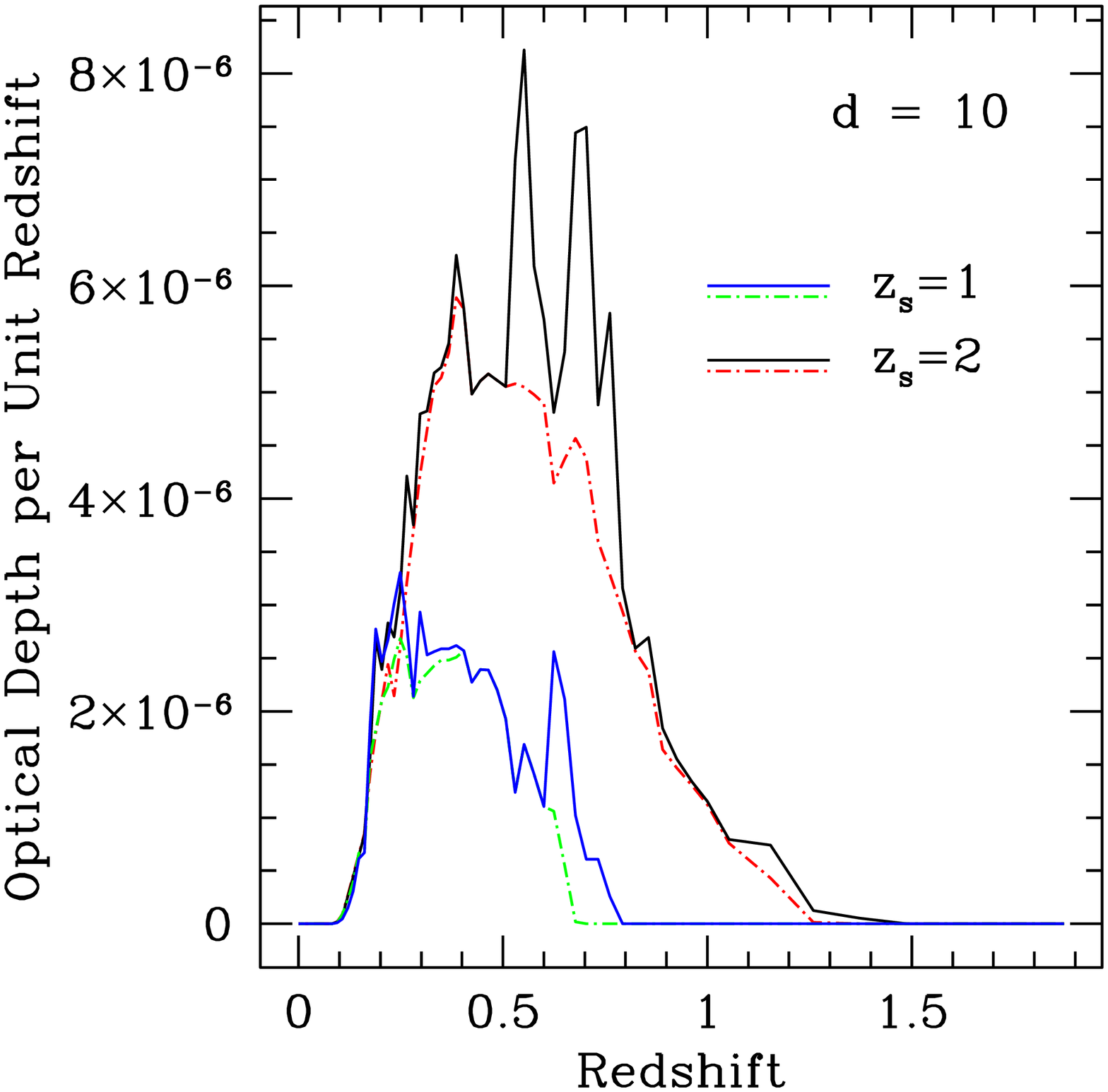}
\caption{The evolution with redshift of the optical depth per unit
  redshift for gravitational arcs with length-to-width ratios equal to
  or larger than $d$. Solid lines include the effect of cluster
  mergers, while dashed lines do not. Upper curves refer to sources with
  $z_\mathrm{s} = 2$ and lower curves to sources with $z_\mathrm{s} = 1$.}
\label{fig:opdepth}
\end{figure}

In Fig.~\ref{fig:opdepth} we show the optical depth per unit redshift,
i.e.~the contribution to the lensing optical depth given by structures
in different redshift bins per unit lens redshift.
In other words, the integrand of the
redshift integral in Eq.~(\ref{eqn:numdepth}) is shown. The
integration over mass is carried out above the mass limit illustrated
in Fig.~\ref{fig:massmax}, i.e.~all halos are included which have
caustic structures sufficiently larger than individual source
galaxies. The two panels are for $d=7.5$ and $d=10$, respectively. Four
curves are shown in each panel, obtained ignoring mergers (smooth,
dashed curves) and taking mergers into account (solid curves). In both
panels, the upper and lower curves refer to source redshifts
$z_\mathrm{s}=2$ and $z_\mathrm{s}=1$, respectively.

The overall trend of the differential optical depth in
Fig.~\ref{fig:opdepth} resembles individual cross sections, i.e.~it
drops to zero as the lenses approach the observer or the sources. 
The dashed (upper)
curve for $z_\mathrm{s}=2$
broadly peaks at redshift $z_p\approx0.4$, slightly larger than the
typical redshift for the peak of the individual cross sections shown
in Fig.~\ref{fig:crsecbis}. This is simply due to the fact that in the
differential optical depth the lensing cross sections are weighted
with the number density of structures within mass bins. It is
interesting to note that the same peak occurs even in the corresponding
solid curve, thus it is not due to dynamical processes in the
cluster lenses, but rather to the combination of the mass evolution
of the lenses
with the particularly high lensing efficiency for clusters at that redshift.

Apart from that, the most remarkable result shown by the solid curves
is that the impact of cluster mergers is important in particular at
moderate and high redshifts, $0.5\lesssim z\lesssim0.8$. The
pronounced peaks in the differential optical depth seen there even
after averaging over the halo sample indicate that cluster mergers can
substantially increase the lensing optical depth of high-redshift
clusters. Above redshift $0.5$, mergers almost double the optical
depth.

Shifting the source plane from $z_\mathrm{s}=2$ to $z_\mathrm{s}=1$
significantly lowers the total optical depth as well as the impact
of cluster
mergers on the optical depth per unit redshift.
The first effect is obviously due to the fact that for lower
source redshift, the redshift interval of high
lensing efficiency narrows. The second effect reflects
that sources at lower redshift miss a significant part of the
lensing haloes' formation history and merger processes related with it.

\begin{figure}[ht]
\begin{center}
  \includegraphics[width=\hsize]{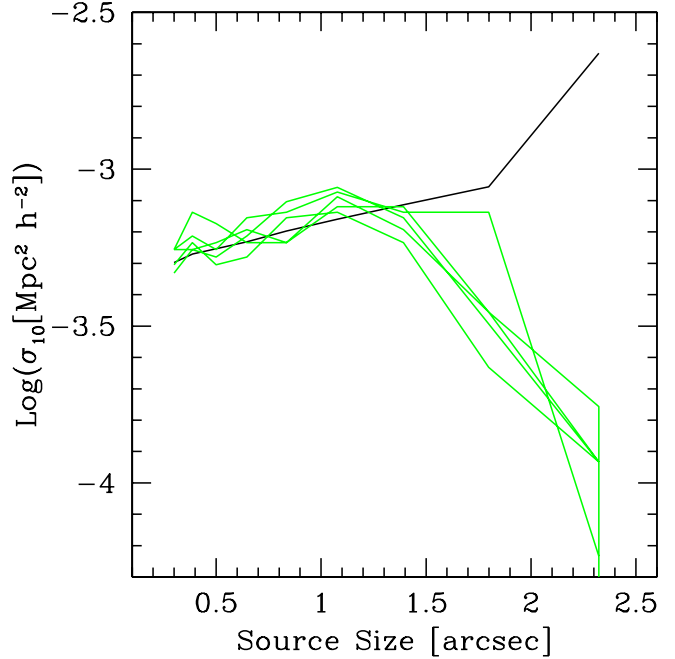}
\end{center}
\caption{Cross section for gravitational arcs with length-to-width ratio
exceeding $d=10$ for a dark-matter halo of $10^{15}\,M_\odot h^{-1}$
as a function of the source size in arc sec. Black and green lines
show the results of the
semi-analytic and of ray-tracing
simulations, respectively, which are
repeated with different random-number seeds for
the ellipticities and the position angles of the sources, as in
Fig.~\ref{fig:crsec}.
The source redshift is $z_\mathrm{s}=1$, and the lens redshift is
$z_\mathrm{l}=0.3$.}
\label{fig:sizes}
\end{figure}

\subsection{Sources Properties}

A full analysis of the effect of various source properties
on the cross sections of individual haloes is well beyond the scope of
this paper.
Moreover, the method we have outlined in the previous subsections is
probably not the ideal tool for that purpose. For example, it would be
of interest
to check the variation of cross sections with the source size.
Fig.~\ref{fig:sizes} shows that numerical and semi-analytic
cross sections agrees (as already shown)
and initially increase with increasing source
size. However, when the sources become too large compared to the
size of the caustic structure, the assumption that
the lensing properties are approximatively constant across the area
of a source fails (see the Appendix for details). Thus, while numerical
cross sections start to decrease because the sources are too large to
be efficiently distorted, the semi-analytic cross section
increases dramatically. We note that, as in the rest of this
paper, dark matter haloes are modelled with NFW density profiles and
elliptically distorted lensing potential.

Nonetheless, some testable effects
can be briefly
addressed here. The first is the influence of the shape
(circular or elliptical) of the sources on the lensing efficiency.
On this ground we
can compare our results with \cite{KE01.2}.

\begin{figure}[ht]
\begin{center}
  \includegraphics[width=\hsize]{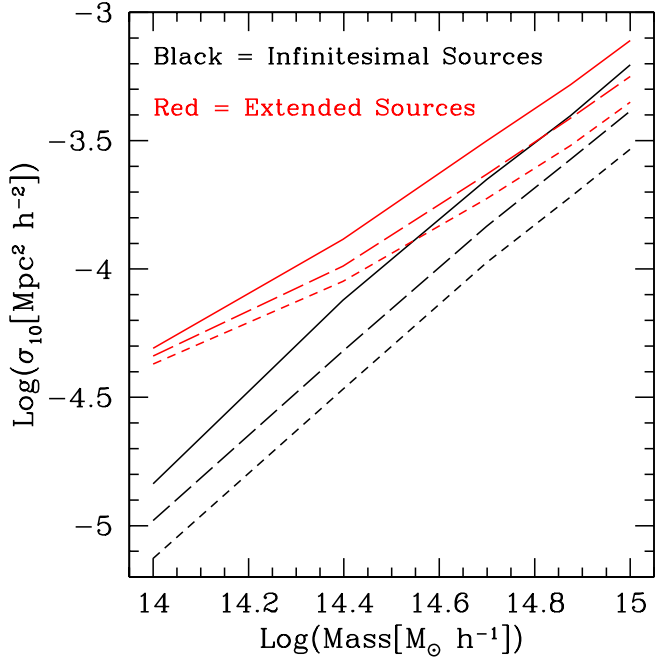}
\end{center}
\caption{Cross section for gravitational arcs with length-to-width ratio
exceeding $d=10$ for several haloes of increasing mass.
Black and red lines refer
to point-like or extended sources with area equal
to that of a circle of radius $0.5''$, respectively.
Short dashed lines indicate circular sources,
long dashed lines refer to sources with random ellipticity drawn from the
interval $[0,0.5]$ $\left( q_s \in [0.5,1] \right)$,
and solid lines refer to sources with ellipticity $0.5$. Sources are at
redshift $z_\mathrm{s}=1$ and the lens redshift is $z_\mathrm{l}=0.3$.}
\label{fig:ellipticity}
\end{figure}

Fig.~\ref{fig:ellipticity}
shows the cross section of several dark-matter
haloes of increasing mass as a function of the source ellipticity, as
explained in the caption.
We see in the Figure that the cross sections for small elliptical
sources exceed those of small circular sources by a factor of
$\approx2$.
This agrees with the findings of
\cite{KE01.2}. It is quite interesting to note that the corresponding
increase is somewhat lower for larger sources.
This is due to the fact that the cross section is on the average higher,
thus the contribution from the source ellipticity is relatively less
important.

Another quite important effect is the change of
the cross section with the source redshift. Investigating this, we will
keep the source size fixed at a radius of $0.5''$, since the angular diameter
distance changes only a little over redshift 1.

\begin{figure}[hb]
\begin{center}
  \includegraphics[width=\hsize]{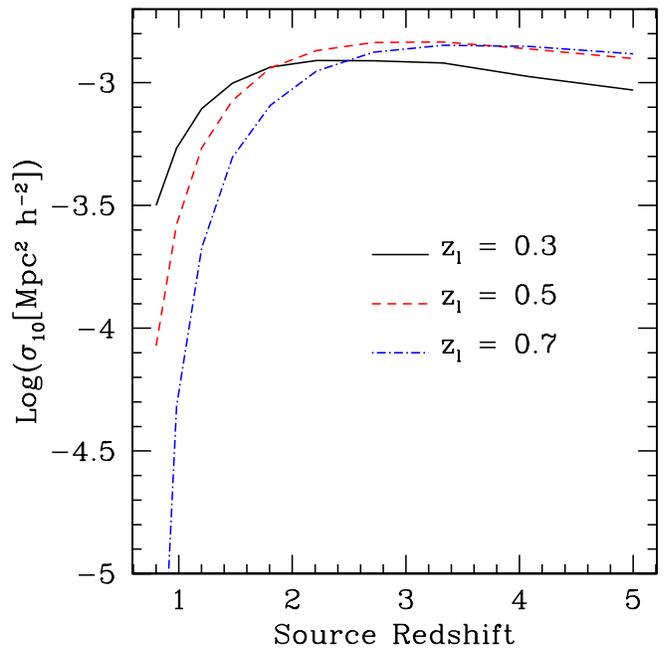}
\end{center}
\caption{Cross section for arcs with length-to-width ratio exceeding
$d=10$ for a dark matter halo of $10^{15}\,M_\odot h^{-1}$ as a function
of the source redshift. Three different lens redshifts are considered, as
labeled in the plot.}
\label{fig:redshift}
\end{figure}

In Fig.~\ref{fig:redshift} we plot the cross section for a
single dark-matter halo of $10^{15} M_{\odot} h^{-1}$ as a function
of the source redshift. We adopt three different lens redshifts, $0.3$,
$0.5$ and
$0.7$ respectively. The low-redshift end
of that plot shows that the closer the lens is to the source, the lower
the cross section is due to the geometrical suppression of the halo's lensing
efficiency (see the discussion in Sect.~5).
The general trend is a rapid increase of the lensing cross section
with increasing lens redshift followed by an almost constant phase and a
slightly decreasing phase. This trend agrees with the general evolution
of the lensing efficiency with the source redshift.

These few examples show how the influence of the source redshift
and intrinsic properties on a cluster's lensing efficiency is significant,
in agreement with earlier work, and a deeper analysis of the
consequences is
certainly needed.
 
\section{Summary and Discussion}

We have described a novel method for semi-analytically calculating
strong-lensing cross sections of galaxy clusters. The method first
approximates the length-to-width ratio of images by the ratio of the
eigenvalues of the Jacobian matrix of the lens mapping. The
requirement that this ratio exceed a fixed threshold defines a stripe
on both sides of the (tangential) caustic curve whose area
approximates the cluster's strong-lensing cross section.

This approach would be valid for infinitesimally small, circular
sources. Extending it for elliptical sources is straightforward using
the elegant technique developed by \cite{KE01.2}. Extended sources can
be taken into account after convolving the eigenvalue ratio with a
suitable window function quantifying the source size. In order to
speed up this convolution, we approximate it by a simple
multiplication.

We have tested this method by detailed comparison with cross sections
obtained from full ray-tracing simulations. We found excellent
agreement within the (considerable) error bars of the ray-tracing
results for a variety of lens masses and lens and source
redshifts. Deviations occur for very weak lenses whose caustics are so
small that the crucial assumption of the eigenvalue ratio's not
changing much across sources is no longer satisfied. In particular,
our tests revealed that cross sections rapidly changing during merger
events are well reproduced by our new method.

We then proceeded to apply the technique to a sample of halos whose
history is described by simulated merger trees. The halos themselves
are modelled as pseudo-elliptical halos with NFW density profile whose
mass is given as a function of redshift by the merger tree. The merger
trees are obtained from a sample of 46 cluster-sized halos numerically
simulated in a cosmological volume. We followed the evolution of the
halos by simulating merger events at times when the merger trees
signal the accretion of a sub-halo with mass comparable to that of the
main halo.

This technique allowed us to study the total optical depth for arc
formation by the simulated cluster sample at a time resolution high
enough for properly following merger events. Comparing the results to
those obtained ignoring mergers, we found that the arc optical depth
produced by clusters with moderated and high redshifts, $z\gtrsim0.5$,
is almost doubled by mergers.

The resuts just outlined may be potentially relevant in view of the
high frenquency of arcs recently detected in
clusters at moderate and high redshifts
\citep{GL03.1,ZA03.1}. For example, \cite{GL03.1} argue
that some physical process must boost the lensing efficiency of
high-redshift clusters which is likely connected with
the dynamics and formation histories of the lensing clusters,
as merger processes are. This conclusion was supported by
\citep{HO05.3} after this paper was first submitted.
\cite{HO05.3} compare the lensing efficiency of matched observed and simulated
(low-redshift) galaxy clusters with a realistic source population
taken from the Hubble Deep Field. They find that real clusters
are a little bit more efficient lenses than simulated ones, and though
the difference is marginally significant, they argue it could be due to
a selection effect connected with cluster mergers. In fact, cluster
mergers increase not only the lensing efficiency, but also the $X$-ray
luminosity \citep{RA02.1}, and the real clusters used by \cite{HO05.3}
are $X$-ray selected. In other words, the observed lensing clusters
could be a biased sub-set of the entire
cluster population.

The method described and developed here reduces computation times for
strong-lensing cross sections by factors of $\sim30$ compared to
ray-tracing simulations. It thus becomes feasible to reliably compute
strong-lensing probabilities describing an evolving cluster population
by halos accreting mass as encoded by simulated merger trees, whose
merging events can be studied at high time resolution. We shall apply
this technique in forthcoming studies with special emphasis on the
importance of cluster evolution for strong lensing.

\section*{Acknowledgements}

This work was supported in part by the Sonderforschungsbereich 439,
``Galaxies in the Young Universe'', of the Deutsche
Forschungsgemeinschaft. The simulations were carried out on the IBM-SP4
machine at the Centro Interuniversitario del Nord-Est per il Calcolo
Elettronico (CINECA, Bologna), with CPU time assigned under an
INAF-CINECA grant.
We wish to thank an anonymous referee for useful remarks that allowed us
to improve the presentation of the work.

\bibliographystyle{aa}
\bibliography{master}

\appendix

\section{Approximate Convolution of a Function With a Step Function on
  the Source Plane}

Consider an arbitrary function $\bar R(\vec y)$ defined on the source
plane. Suppose we wish to convolve $\bar R(\vec y)$ with another
function $g(\vec y)$, defined as in Eq.~(\ref{eqn:step}). The
convolution is
\begin{equation}
  h(\vec y)=(\bar R*g)(\vec y)=
  \int_{\mathbb{R}^2}\,\bar R(\vec z)\,g(\vec y-\vec z)\d^2z\;.
\end{equation}
Without loss of generality, any given point $\vec y$ on the source
plane can be chosen as the coordinate origin. This means that
$y\equiv0$, hence
\begin{equation}
  h(0)=(\bar R*g)(0)=
  \int_{\mathbb{R}^2}\,\bar R(\vec z)\,g(\vec z)\d^2z\;.
\end{equation}
Now, we choose a position $\vec u$ on the \emph{lens plane} such that
$\vec z=\vec z(\vec u)$. Applying the lens mapping to the convolution
above we obtain
\begin{equation}
  h(0)=\int_{\mathbb{R}^2}\,R(\vec u)\,g(\vec u)\,
  \frac{\d^2u}{|\mu(\vec u)|}\;.
\end{equation}
We assume that $\bar R(\vec y)$ does not vary much across a
source. This assumption is satisfied in almost all interesting cases,
except when the sources are at high redshift and the lens is close to
them. In that case the critical curves are very small, thus the
typical scale on which the lensing properties vary may be comparable
with the angular extent of a source. However, even in that case the
results of our method remain good, in particular in view of the
substantial scatter in the ray-tracing results.

Within this assumption, we can expand the function $R(\vec x)$ into a
Taylor series around zero, obtaining (summing over repeated
indices)
\begin{eqnarray}
  h(0)&\approx&
  R(0)\int_{\mathbb{R}^2}\,g(\vec u)\,\frac{\d^2u}{|\mu(\vec u)|}+
  \frac{\partial R(0)}{\partial u_i}
  \int_{\mathbb{R}^2}u_ig(\vec u)
  \frac{\d^2 u}{|\mu(\vec u)|}\nonumber\\
  &+&\frac{1}{2}\frac{\partial^2R(0)}{\partial u_i\partial u_j}
  \int_{\mathbb{R}^2}u_iu_jg(\vec u)\frac{\d^2u}{|\mu(\vec u)|}\;.
\label{eqn:a4}
\end{eqnarray}
The first integral is unity by normalisation, and the second vanishes
because of the symmetry of $g(\vec x)$. Thus, Eq.~(\ref{eqn:a4})
reduces to
\begin{eqnarray}
  h(\vec y)&\approx&R(0)+
  \frac{1}{2}\frac{\partial^2R(0)}{\partial u_i\partial u_j}
  \int_{\mathbb{R}^2}u_iu_jg(\vec u)
  \frac{\d^2 u}{|\mu(\vec u)|}\nonumber\\
  &\equiv&R(0)+
  \frac{1}{2}\frac{\partial^2R(0)}{\partial u_i\partial u_j}
  \Omega_{ij}\;.
\label{eqn:expansion}
\end{eqnarray}
Thus, we have to carry out the three integrals $\Omega_{11}$,
$\Omega_{22}$ and $\Omega_{12}=\Omega_{21}$. We show the explicit
calculation only for the first, as the others are quite similar:
\begin{equation}
  \Omega_{11}\equiv
  \int_{\mathbb{R}^2}u_1^2g(\vec u)\frac{\d^2u}{|\mu(\vec u)|}=
  \frac{1}{\pi\zeta^2}\int_{D}u_1^2\frac{\d^2 u}{|\mu(\vec u)|}\;,
\label{eqn:omega}
\end{equation}
where $D$ is the set of all positions $\vec x$ on the lens plane where
$\vec x^T\Gamma\vec x\le1$. The matrix $\Gamma$ defines the shape of
the image formed from the source and can be written as
$\Gamma=\mathcal{A}^T\mathcal{B}\mathcal{A}=
\mathcal{A}^T\mathcal{A}/\zeta^2$. Obviously, the eigenvalues of
$\Gamma$ are $\lambda_\mathrm{t}^2/\zeta^2$ and
$\lambda_\mathrm{r}^2/\zeta^2$. We can now rotate into a reference
frame in which $\mathcal{A}$ and thus also $\Gamma$ are diagonal. This
is achieved by a rotating about an angle
\begin{equation}
  \varphi=\frac{1}{2}\arctan\left(\frac{\gamma_2}{\gamma_1}\right)\;,
\end{equation}
where $\gamma_1$ and $\gamma_2$ are the two shear components at the
origin. This rotation is described by the orthogonal matrix
\begin{equation}
  \mathcal{R}=\left(\begin{array}{cc}
    \cos\varphi & -\sin\varphi\\
    \sin\varphi &  \cos\varphi\\
  \end{array}\right)\;.
\end{equation}
With $\vec v=R\vec u$, this rotation, transforms Eq.~(\ref{eqn:omega})
into
\begin{eqnarray}
  \Omega_{11}&=&\frac{1}{\pi\zeta^2}
  \int_{D}(v_1^2\cos^2\varphi+v_2^2\sin^2\varphi)
  \frac{\d^2v}{|\mu(\vec v)|}\nonumber\\
  &+&\int_{D}(2v_1v_2\cos\varphi\sin\varphi)
  \frac{\d^2 v}{|\mu(\vec v)|}\nonumber\\
  &=&\frac{\cos^2\varphi}{\pi\zeta^2}\int_{D}
  v_1^2\frac{\d^2v}{|\mu(\vec v)|}+\frac{\sin^2\varphi}{\pi\zeta^2}
  \int_{D}v_2^2\frac{\d^2 v}{|\mu(\vec v)|}\;.
\label{eqn:omega2}
\end{eqnarray}
In the new coordinate system, $D$ is the set of all positions $\vec v$
of the lens plane where
\begin{equation}
  \frac{\lambda_\mathrm{t}^2}{\zeta^2}v_1^2+
  \frac{\lambda_\mathrm{r}^2}{\zeta^2}v_2^2\le1\;.
\end{equation}
Now we can introduce polar elliptical coordinates $(\rho,\theta)$ by
\begin{equation}
  v_1=\frac{\zeta}{|\lambda_\mathrm{t}|}\rho\cos\theta\;,\quad
  v_2=\frac{\zeta}{|\lambda_\mathrm{r}|}\rho\sin\theta\;,
\end{equation}
in terms of which Eq.~(\ref{eqn:omega2}) becomes
\begin{eqnarray}
  \Omega_{11}&=&\frac{\zeta^2\cos^2\varphi}{\pi}
  \int_0^{2\pi}\int_0^1\rho^3\cos^2\theta
  \frac{\d\rho\d\theta}{\lambda_\mathrm{t}^2(\rho,\theta)}\nonumber\\
  &+&\frac{\zeta^2\sin^2\varphi}{\pi}
  \int_0^{2\pi}\int_0^1\rho^3\sin^2\theta
  \frac{\d\rho\d\theta}{\lambda_\mathrm{r}^2(\rho,\theta)}\;.
\end{eqnarray}

As a second approximation, we shall assume that the tangential and
radial eigenvalues do not vary much across a single image, so we can
replace the eigenvalues by their mean values across the image. Then,
\begin{eqnarray}
  \Omega_{11}&=&
  \frac{\zeta^2\cos^2\varphi}
       {\pi\langle\lambda^2_\mathrm{t}\rangle}
  \int_0^{2\pi}\int_0^1\rho^3\cos^2\theta\d\rho\d\theta\nonumber\\
  &+&\frac{\zeta^2\sin^2\varphi}
          {\pi\langle\lambda_\mathrm{r}^2\rangle}
  \int_0^{2\pi}\int_0^1\rho^3\sin^2\theta\d\rho\d\theta\nonumber\\
  &=&\frac{\zeta^2}{4}\left(
  \frac{\cos^2\varphi}{\langle\lambda_\mathrm{t}^2\rangle}+
  \frac{\sin^2\varphi}{\langle\lambda_\mathrm{r}^2\rangle}
  \right)\;.
\end{eqnarray}
Similarly, we find that
\begin{equation}
  \Omega_{22}=\frac{\zeta^2}{4}\left(
  \frac{\cos^2\varphi}{\langle\lambda_\mathrm{r}^2\rangle}+
  \frac{\sin^2\varphi}{\langle\lambda_\mathrm{t}^2\rangle}
  \right)
\end{equation}
and
\begin{equation}
  \Omega_{12}=\frac{\zeta^2\sin\varphi\cos\varphi}{4}\left(
  \frac{1}{\langle\lambda_\mathrm{r}^2\rangle}-
  \frac{1}{\langle\lambda_\mathrm{t}^2\rangle}
  \right)
\end{equation}
Substituting into Eq.~(\ref{eqn:expansion}), we obtain
\begin{eqnarray}
  h(0)&\approx&R(0)+\frac{1}{2}\frac{\partial^2R(0)}{\partial x_1^2}
  \left(\frac{\zeta^2\cos^2\varphi}
             {4\langle\lambda^2_\mathrm{t}\rangle}+
        \frac{\zeta^2\sin^2\varphi}
             {4\langle\lambda^2_\mathrm{r}\rangle}
  \right)\nonumber\\
  &+&\frac{1}{2}\frac{\partial^2R(0)}{\partial x_2^2}
  \left(\frac{\zeta^2\cos^2\varphi}
             {4\langle\lambda^2_\mathrm{r}\rangle}+
        \frac{\zeta^2\sin^2\varphi}
             {4\langle\lambda^2_\mathrm{t}\rangle}\right)\nonumber \\
  &+&\frac{\partial^2R(0)}{\partial x_1\partial x_2}
  \left(\frac{\zeta^2\cos\varphi\sin\varphi}
             {4\langle\lambda^2_\mathrm{r}\rangle}-
        \frac{\zeta^2\cos\varphi\sin\varphi}
             {4\langle\lambda^2_\mathrm{t}\rangle}\right)\;.
\label{eqn:convolution}
\end{eqnarray}
Within the framework of our approximations, we can thus replace the
value of the function $R$ at a point of the lens plane with its
convolution on the source plane at the corresponding point,
represented by Eq.~(\ref{eqn:convolution}). In this way, we account
for finite source sizes.

\end{document}